\documentclass[
superscriptaddress,
amsmath,amssymb,
 preprintnumbers,
twocolumn,
pra,
aps,
floatfix,
]{revtex4-1}

\usepackage{graphicx}
\usepackage{dcolumn}
\usepackage{bm}
\usepackage{chngcntr} 
\usepackage{hyperref}
\usepackage[mathlines]{lineno}
\usepackage{physics}
\usepackage{tikz}
\usetikzlibrary{quantikz}
\usepackage{algpseudocode}
\usepackage{algorithm}
\usepackage{lineno}
\usepackage{booktabs}
\usepackage{adjustbox}

\newcommand{\vsp}[1][0.10cm]{ \vspace{#1}}

\begin{document}

\title{An Alternative Approach to  Quantum Imaginary Time Evolution}
% Alphabetical, last name
% Please correct the affiliations
\author{Pejman Jouzdani}
% Authors- Affiliation/contacts
\email{Corresponding author, email: jouzdanip@fusion.gat.com}
\affiliation{General Atomics, 3550 General Atomics Ct, San Diego, CA 92121, USA}
%
% Authors- Affiliation/contacts
\author{Calvin W. Johnson}
\affiliation{San Diego State University, 5500 Campanile Drive, San Diego, CA 92182, USA}
%
% Authors- Affiliation/contacts
\author{Eduardo R. Mucciolo}
\affiliation{Department of Physics, University of Central Florida, Orlando, FL 32816, USA}
%
% Authors- Affiliation/contacts
\author{Ionel Stetcu}
\affiliation{Theoretical Division, Los Alamos National Laboratory, Los Alamos, NM 87545, USA}

\date{\today}

\clearpage

%%%%%%%%%%%%%%%%%%%%%%
% \input{paper-abstract}
%%%%%%%%%%%%%%%%%%%%%%
\begin{abstract}
There is increasing interest in quantum algorithms that are based on the imaginary-time evolution (ITE), a successful classical numerical approach to obtain ground states. However, 
most of the proposals so far require heavy post-processing 
computational steps on a classical computer, such as solving linear equations. Here we 
provide an alternative approach to implement ITE.
A key feature in our
approach is the use of an orthogonal basis set: the propagated state is efficiently expressed in terms of orthogonal basis states at every step of the evolution. We argue that the number of basis states needed at those steps to achieve an accurate solution can be kept of the order of $n$, the number of qubits, by controlling the precision (number of significant digits) and the imaginary-time increment. The number of quantum gates per imaginary-time step is estimated to be polynomial in $n$.
Additionally, while in many QAs 
the locality of the Hamiltonian is a key assumption, in our algorithm this restriction is not required. This characteristic of our algorithm renders it useful for studying highly nonlocal systems, such as the occupation-representation nuclear shell model. We illustrate our algorithm through numerical implementation on an IBM quantum simulator.
\end{abstract}
% \clearpage

% meta keywords for search and tagging in the search engine

\clearpage

% Base constructor
\maketitle

\preprint{LA-UR-22-26600}  % Los Alamos Unlimited Release number
% Sections
%%%%% opening intro
%%%%%%%{Opening intro}
\section{Introduction}
\label{sec:introduction}

% QC, what is it in our view
Gate-based quantum computers are promising platforms to 
realize a wide range of arbitrary unitary matrices not 
accessible via classical computers due mostly to 
memory-size restrictions. The fundamental idea is that any unitary 
operator $U$ can be decomposed as a series of products of 
smaller unitary operations, known as quantum gates, which can 
be directly implemented on a quantum computer. A unitary operator $U$ is then equivalent to a quantum circuit constructed 
from those quantum gates.
% PQC definition
An approach being considered is a parametric quantum circuit (PQC), where a series of adjustable 
(parametric)
quantum gates results in a unitary matrix  with parameters 
${\bf y}$, namely, $U({\bf y})$.
\par

% QC, what to do and how
Given a quantum computer, the natural question is what can 
be computed and how to instruct the machine, that is, which 
quantum algorithm to use. A major class of quantum algorithms for 
state preparation is based on the variational principle 
\cite{Peruzzo2014}. The idea is to minimize a cost function 
such that $U({\bf y}) \ket{r}$ closely approximates a target 
state $\ket{\psi}$ for a suitable initial state $\ket{r}$. Algorithms in this class are generally 
referred to as a variational quantum algorithms (VQAs) 
\cite{Cerezo2020VariationalAlgorithms}. In the current era of 
noisy-intermediate scale quantum (NISQ) processors 
\cite{Preskill2018}, VQAs are considered less costly 
in terms of circuit depth.
However, it is still unclear how to choose and design 
the arrangement of PQC gates in general and how to avoid the appearance of barren plateaus during the optimization process \cite{Larocca2022, Sack2022}.   

There is growing interest in quantum algorithms that 
are based instead on classical imaginary-time evolution (ITE).
%%%% idea
The basic idea is to replace the real time $t$ with an 
imaginary time $\tau = it$, so that the initial state is projected onto the ground state with exponential accuracy.
%%%% Impossible to implement
Since ITE  is 
a non-unitary evolution, direct implementation on a 
quantum computer is impossible.
%%%% ITE -> VITE;
Recently, the variational principle was used to optimize 
a PQC \cite{McArdle2019, Jones2019b} such that the ITE 
is replicated by a unitary evolution of the quantum circuit. 
%%%% ITE -> QITE; 
In an other attempt \cite{Motta2019DeterminingEvolution}, a 
quantum version of ITE (named QITE) was introduced where quantum gates are chosen 
at every time step. 
%%%% ITE -> PITE
Others approaches utilize ancillary qubits where the algorithm operates on the 
probability of measuring a correct 
outcome \cite{Kosugi2021, Turro2022}. 
%%%% ITE -> QITE compression of circuit.
Efforts to further compress the quantum circuit have been reported as well 
\cite{Lin2021, Nishi2021ImplementationApproximation}.

%%%% Applications
Applications of ITE in physics and chemistry are many, from directly finding the ground state of a target Hamiltonian to the computation of the energy spectra, which can be achieved by 
quantum and Krylov subspace \cite{Motta2019DeterminingEvolution} expansions. The implementation of ITE on quantum computers would also allow the computation of Green’s functions \cite{Wecker2015}, the biggest bottleneck in dynamical mean-field theory calculations \cite{Kotliar2006}. This possibility has motivated the pursue of hybrid quantum-classical simulations of correlated materials
\cite{Sakurai2022HybridFunctions, Bauer2016, Kreula2016}.

In this paper, we propose a modified implementation of ITE. Our proposal is inspired by the QITE of Ref.  \cite{Motta2019DeterminingEvolution}, but with a major modification that circumvents the main challenge of that method. In QITE, a quantum circuit $U_{\tau}$ is used to connect an initial state $\ket{\textbf{0}}$ to the final state $\ket{\psi_{\tau}}$ that would have resulted from an imaginary time evolution, namely, $\ket{\psi_{\tau}} =U_{\tau}\ket{\textbf{0}}$. In other words, the imaginary time evolution is replaced by an equivalent unitary operation.
%%%%
To obtain the quantum state at a time $\tau+\delta$, $\ket{\psi_{\tau+\delta}}$, an additional quantum circuit $U_{\delta}$ must be found and applied to $U_{\tau}\ket{\textbf{0}}$, that is, 
\begin{eqnarray}
\ket{\psi_{\tau+\delta}} & = & U_{\delta} \ket{\psi_\tau}
\nonumber \\
& = & U_{\delta}\, U_{\tau} \ket{\textbf{0}}.
\label{eq-QITE-1}
\end{eqnarray}
The challenge is to identify the gates involved in $U_{\delta}$ and the corresponding parameters.

The authors of Ref. \cite{Motta2019DeterminingEvolution} propose the  following solution to the above challenge:
Consider the difference between the two states  
$\ket{d\psi} = \ket{\psi_{\tau+\delta}} - \ket{\psi_{\tau}}$, which is known in terms of the Hamiltonian and $\ket{\psi_{\tau}}$.
Assume a unitary $U_\delta$ comprising of a set of parametric gates. 
These gates are limited in number due to a bounded correlation length. Once $U_\delta$ is chosen, $ U_{\delta}
\ket{\psi_\tau} - \ket{\psi_\tau} $ is expanded and set equal to 
$\ket{d\psi}$. Ultimately, by solving 
a set of linear equations on a classical computer the gate parameters are found. A major drawback of this method is that the matrix dimensions of this linear system of equations are not in general bounded and can easily exceed a classical computer capacity. 

In addition, in Ref. \cite{Motta2019DeterminingEvolution} 
$\ket{d\psi}$ is expressed in terms of a set of non-orthogonal and incomplete 
eigenvectors, making the solution of the linear system of equations difficult. A simple solution to avoid non-orthogonality is to append $U_{\delta}$ differently:
Consider swapping $U_\tau$ and $U_{\delta}$ in Eq. (\ref{eq-QITE-1}), that is,
\begin{eqnarray}
\ket{\psi_{\tau+\delta}} 
&=& 
U_{\tau}\, U_{\delta} \ket{\textbf{0}}
\nonumber \\
&=& U_{\tau+\delta} \ket{\textbf{0}}.
\label{eq-MQITE-1}
\end{eqnarray}
By properly choosing the gates in $U_{\delta}$, 
the expansion of $U_{\delta} \ket{\textbf{0}}$ 
can be found to be a sum of {\it orthogonal states}, which, consequently, results in an expansion of $\ket{\psi_{\tau+\delta}} $ in terms of some orthogonal basis. 
More specifically, this is accomplished by considering 
$U_{\delta}= e^{iy_j P_j}$, where $P_j$
is a Pauli string operator, and by starting with a 
state $\ket{\textbf{0}}$ that is a product of 
individual qubit states. 

In the following sections, we outline the algorithm that
identifies the 
gate parameter set ${\bf y} = \{y_j\}$ and the corresponding set 
$\{P_j\}$ of operators in $U_\delta$ at every step 
of the evolution. We also show that, in this approach, there is no need to solve linear equations off-line or on a classical computer. Additionally, the necessary gates are naturally obtained and there is no 
requirement of locality of the Hamiltonian or bounded correlation length. 

Our algorithm is a modified  version of QITE, and thus we call it MQITE. Our algorithm does make a few assumptions. It assumes that it is possible to measure an
$\eta$ number of state vector components at every time step. 
The number $\eta$ is a function of $\delta$, the imaginary-time increment, and $\epsilon$, the desired precision in the measurement. For most examples we studied,
$\eta $ is of order of $n$ or $n^2$, where $n$ is the number of qubits.

%%%% Not on NISQ
Generally implementing an algorithm that replicates an ITE on a
quantum computer requires deep quantum circuits. In this paper, we assume that the quantum computer is noise-free and the decoherence time is long enough to allow for long-depth circuits, which may not yet be achievable in most NISQ device implementations at the moment.  

% Except perhaps for certain problems where a compact representation of gates arrangement exists \cite{Nishi2021ImplementationApproximation}

%%%%%%%%%%%%%%%%%%%%%%%%%%%%%%%%
%%%%%%%%%%%%%%%%%%%%%%%%%%%%%%%%
%%%%%%%%%%%%%%%%%%%%%%%%%%%%%%%%
%%%%%%%%%%%%%%%%%%%%%%%%%%%%%%%%

Definitions and notations used in the paper are outlined in Sec. \ref{sec-background}. 
Readers familiar with those aspects may skip to Sec. \ref{sec-MQITE}
for a detailed presentation of the MQITE and its steps. In Sec. \ref{sec-numerical}, numerical examples are provided. We show the efficiency of the algorithm 
for a range of problems, including a classical spin problems on arbitrary graphs, the so-called Max-Cut problem \cite{LucasA2014}, as well as the one-dimensional transverse-field Ising model (TFIM), 
and finally the configuration-interaction nuclear shell model in occupation space. A summary and an outlook are provided in Sec. \ref{sec-Conclusion}. A subroutine for the decomposition
of the many-qubit gate used in the algorithm 
is provided in Appendix \ref{sec:Decomposition}.
% appendix on phase calculation
A sub-algorithm to compute the number of state vector components $\eta$ is presented in Appendix \ref{sec:MeasurementCircuit}.

\section{Background}
\label{sec-background}

%%%%%%%
\subsection{Definitions}
\label{sec-definitions}
%%% Notation
In the following, we express an operator with an uppercase letter and a scalar with a lowercase one. 
%%%%%%%% Pauli
A Pauli string operator $P$ is a tensor product of Pauli operators 
$
\{X,
Y,
Z
\}
$
and the identity $I$. Specifically,
\begin{eqnarray}
P_j = O^{(j)}_1 \otimes \cdots \otimes O_n^{(j)},
\label{eq-Pauli-string}
\end{eqnarray} 
where $n$ is the number of qubits and $O^{(j)} \in 
\{X,
Y,
Z, 
I
\}
$. 
A Pauli string operator is Hermitian and unitary, that is, $P=P^\dagger$ and   $ (P)^2 =I_1 \otimes \cdots \otimes I_n= I^{\otimes n}$. We may use \emph{Pauli string operator} and 
\emph{Pauli string} interchangeably.

%%% Hamiltonian
A Hamiltonian is considered to be a sum of Pauli string operators and thus defined by its set of Pauli strings and the corresponding coefficients. 
Thus a Hamiltonian $H$, 
defined over $n$ qubits, is written 
as a weighted sum of all Pauli string operators (the identity included),
\begin{eqnarray}
H = \sum_k w_k Q_k,
\end{eqnarray}
where $Q_k$ is a Pauli string and the weight $w_k$ is a real number.
\par

%%%%%%% a unitary 
A unitary operator $e^{i y_j P_j}$ can be expressed as
\begin{eqnarray}
e^{i y_j P_j} = 
\cos{y_j} \,I^{\otimes n} + i 
\sin{y_j} \, P_j,
\end{eqnarray}
where $y$ is a real parameter. Throughout this paper, $i$ is reserved for the imaginary number 
$i=\sqrt{-1}$.
%%%%%%% decomposition
Each unitary $e^{i y_j P_j}$ can be shown to be implemented by $O(2n)$ one- and two-qubit gates on a fully-connected quantum device \ref{sec:Decomposition}.

%%%%%%% Generic Unitary U
A generic unitary operator acting on $n$ qubits is 
\begin{eqnarray}
U&=& U({\bf y}) \nonumber \\ 
&=&
e^{i\sum_{j=1}^{4^n-1} \,\, y_j P_j} \,\, 
;\,
P_j\ne I^{\otimes N},
\label{eq-tb-U}
\end{eqnarray}
where the exclusion of the identity ensures $\mbox{det}(U)=1$. 

%%%% Trotterization
On gate-based quantum computers, any unitary operation is decomposed into smaller unitary gates which operate  on only one or two qubits, at most.
A usual way to decompose an arbitrary $U$ is by performing trotterization:
\begin{eqnarray}
U &=& \left( e^{i\sum_{j=1}^{4^n-1} \delta_j P_j} \right)^{m_\tau} 
\nonumber \\
&\approx& \left(\prod_{j=1}^{n_u} e^{i \delta_j P_j} \right)^{m_\tau,},
\label{eq-trotter}
\end{eqnarray}
where $n_u$ is at most $4^n-1$, $m_\tau$ is the assumed number of trotterization steps, and $\delta_j = y_j/m_\tau$. Notice the approximate nature of Eq. (\ref{eq-trotter}), as the Pauli strings $P_j$ do not necessarily commute. The error incurred in the approximation is $O(m_\tau\, \delta_j^2)$.

\subsection{Quantum Imaginary Time Evolution}
\label{sec-iTEandQITE}
Through an ITE, a quantum mechanical system evolves toward its ground state since $\ket{\psi_{\tau} } = \frac{e^{-\tau H} \ket{r}}{\vert\vert e^{-\tau H} \ket{r}\vert\vert} $ approaches the true ground state of the Hamiltonian $H$ in the limit  of $\tau \rightarrow \infty$, provided that the ground state has a nonzero overlap with the initial state $\ket{r}$. 
%%%

QITE, introduced in Ref. \cite{Motta2019DeterminingEvolution},
is an attempt to replicate the ITE state at every time increment in the evolution process. Explicitly, one is interested in finding the unitary $e^{i \sum_j y_j  P_j}$ such that 
\begin{eqnarray}
\frac{
e^{-\delta\, H} 
\ket{ \psi_{\tau} } }
{\vert\vert e^{-\delta\, H} 
\ket{ \psi_{\tau} }
\vert\vert}
&=& 
e^{i \sum_j y_j  P_j}
\ket{\psi_\tau}
\label{eq-qite}
\end{eqnarray}
for  a small $\delta$.
In the case when $H=Q$ is a single Pauli string, and after expanding both sides of this equation, one arrives at a set of 
linear equations which is directly associated to the correlations 
$\bra{\psi_\tau} 
P_jP_{j^\prime}
\ket{\psi_\tau}$.
Upon solving the system of equations, the parameters $y_j$ are determined. 
However, the linear equations do not specify the set of $\{P_j\}$
in Eq. (\ref{eq-qite}). These operators are postulated based on the relevant correlation length in the problem, and recently suggested to be determined by combinatorial analysis \cite{Nishi2021ImplementationApproximation}
and other considerations \cite{tsukiyama2011medium}.
The equations for a generic set of $\{P_j\}$ may have a non-zero null-space, and may require a generalized inversion method or an iterative algorithm such as conjugate gradient in order to be solved.

This shortcoming of QITE is a direct consequence of the expansion 
of $\ket{\psi_{\tau+\delta} }$ in terms of an incomplete and non-orthogonal basis set, which in turn is due to the correlations amongst operators. To understand this point, consider the first-order expansion 
\begin{eqnarray}
e^{i \sum_j y_j  P_j} 
\ket{\psi_\tau} 
 & \approx & 
 \ket{\psi_{\tau}} + \sum_j y_j \ket{\phi_j},
\end{eqnarray}
where $\ket{\phi_j}=i P_j\ket{\psi_\tau}$ and, in general, $\bra{\phi_j}\ket{\phi_{j^\prime}} \ne 0$. A similar situation arises in classical techniques based on neural networks that have been recently used to obtain the ground state of 
many-body quantum systems \cite{Carleo2017SolvingNetworks}. Within that context, the non-orthogonality was suggested to be associated with the geometry of the Hilbert space \cite{ChaePark2020}, but can in fact be traced back to the stochastic reconfiguration method of Sorella {\it et al}. \cite{Sorella2007}.

One of the purported advantages of using a quantum computer and 
quantum algorithms is to avoid complications that exist in classical techniques such as the one noted. In this paper, with a simple rearrangement of factors, we arrive at an incomplete but orthogonal basis set $\{\ket{\xi_j}\}$. Due to this orthogonality, no correlations exist and the parameter values come directly from measurements in the quantum circuit.

\section{Modified Quantum Imaginary-time Evolution (MQITE) Algorithm}
\label{sec-MQITE}
\subsection{The Main Idea}
\label{sec-MainIdea}
Consider an alternative approach where
\begin{eqnarray}
\frac{
e^{-\delta Q}
\ket{\psi_\tau}
}
{
\vert\vert
e^{-\delta  Q}
\ket{\psi_\tau}
\vert\vert
} &=& 
U_{\tau}\, 
U_{\delta}
\ket{\textbf{0}}
\nonumber \\ 
&=& 
U_{\tau} \left[
e^{i \sum_j y_j  P_j}\right]
\ket{\textbf{0}}.
\label{eq-mqite-idea}
\end{eqnarray}
As before, $U_{\tau}$ is a known unitary that has been identified in the previous steps of the algorithm such that $\ket{\psi_\tau} = U_{\tau}\ket{\textbf{0}}$.
In contrast to Eq. (\ref{eq-qite}), 
$U_{\delta}$ appears to the right of $U_{\tau}$.
$Q$ is a single Pauli string operator. 
The initial state
$\ket{\textbf{0}} = \ket{0_1, \cdots 0_n}$
is a product state. Upon expansion of the above,
\begin{eqnarray}
\ket{\psi_{\tau+\delta} } 
&\approx&
 \ket{\psi_\tau} + i \sum_j y_j\,  U_{\tau}\, P_j\ket{\textbf{0}}
\nonumber \\
& \approx & 
\ket{\psi_{\tau}} + \sum_j y_j\, U_{\tau} \ket{j}
\nonumber \\
& \approx & 
\ket{\xi_0(\tau)} + \sum_j y_j \ket{\xi_j(\tau)},
\label{eq-MQITE-ortho}
\end{eqnarray}
where $ \ket{j} = i P_j\ket{\textbf{0}}$ is a computational basis state (by construction) and $\ket{\xi_j(\tau)} = U_{\tau} \ket{j}$. The notable difference from the original QITE formulation is that 
% now, 
$\bra{\xi_{j'}(\tau)}\ket{\xi_j(\tau)} = \bra{j'}\ket{j} = \delta_{j'j}$
if we choose a set $\{P_j\}$ of Pauli string 
operators in Eq. (\ref{eq-mqite-idea}) such that the generated bit strings $\{\ket{j}\}$ are distinct. 
As a result,  $\{\ket{\xi_j(\tau)}\}$ is an orthogonal basis set. 

% Suppose the set $\{j\}$ is known to us, 
% then  the operators $\{P_j\}$ are determined straightforwardly. 
% In fact, we can choose two operators for each $\ket{j}$ such that 
% $P^{(r)}_j\ket{\textbf{0}}=-i\ket{j}$ and $P^{(i)}_j\ket{\textbf{0}}=\ket{j}$. This allows us to keep the gate parameter $y_j$ real. 
% We return to this point and how to identify the set $\{j\}$ below in more details.

At first glance, Eq. (\ref{eq-MQITE-ortho}) 
may suggest that there could be an exponential number of components to be evaluated.
This is unlikely from a physical standpoint: Consider evolving the initial 
state $\ket{\textbf{0}}$ by a propagator $U_\tau$, then 
perturb the system by $Q$ (a Pauli string) that could be interpreted physically as an external field, and finally evolve back to time $\tau=0$ with the propagator $U_{-\tau} = U^{\dagger}_{\tau}$. At this time, the state should not significantly differ from 
$\ket{\textbf{0}}$, that is, the projection of 
the state  outside of $\ket{\textbf{0}}$ should be small. This means, in an
active picture, the state $\ket{\psi_{\tau+\delta}}$ differs from $\ket{\xi_0(\tau)} = \ket{\psi_{\tau}}$ by some limited number of components, captured by projecting onto some limited eigenvectors  $\{ \ket{\xi_j(\tau)} = U_{\tau} \ket{j}\}$, with $j\ne0$.

Based on this physical consideration, expanding  
Eq. (\ref{eq-mqite-idea}) 
and multiplying both sides by $U_{\tau}^\dagger$ we arrive at
\begin{eqnarray}
\frac{ 
\ket{\textbf{0}}
-\delta 
U_{\tau}^\dagger Q\, U_{\tau} \ket{\textbf{0}}  
}
{n_\tau} 
= 
\ket{\textbf{0}}+i\sum_j y_j\, P_j\ket{\textbf{0}}.
\nonumber \\
\label{eq-MQITE-main-circuit0}
\end{eqnarray}
Here, $n_\tau$ is a normalization factor,
\begin{eqnarray}
n_\tau = \sqrt{
1-2 \delta 
\bra{\textbf{0}} 
U_{\tau}^\dagger\, Q\, U_{\tau} 
\ket{\textbf{0}} +
\delta^2 
}.
\label{eq-norm}
\end{eqnarray}
After a Taylor expansion of $n^{-1}_\tau$ up to the order $\delta$, and inserting it in Eq. (\ref{eq-MQITE-main-circuit0}), we arrive at 
\begin{eqnarray}
-i\sum_j y_j\, P_j\ket{\textbf{0}}
&=& 
\delta \sum_{j\ne 0}
\ket{j} \, \, \bra{j} U_{\tau}^\dagger\, Q\, U_{\tau} \ket{\textbf{0}}
\nonumber \\
&=&
\delta \sum_{j\ne 0}
c_j \, \ket{j}
\label{eq-MQITE-main-circuit}
\end{eqnarray}
Notice that the state $U_{\tau}^\dagger\, Q\, U_{\tau}\ket{\textbf{0}}$
can be prepared on a quantum computer since $U$ and $Q$ are unitary.
By preparing and executing this circuit a $\chi$ number of times and measuring qubits in the computational basis, a distribution over the bit-strings $\ket{j=0}$ to  $\ket{j=2^n-1}$ is obtained. 

Essentially, from the right-hand side (RHS) of Eq. (\ref{eq-MQITE-main-circuit}), for every 
bit string $\ket{j}$, observed after measurements in a quantum computer,
a Pauli string operator $P_j$ can be construct to satisfy the left-hand-side (LHS) of the same equation. $P_j$ generates $\ket{j}$ upon acting on $\ket{\textbf{0}}$.
% $\ket{j} = i P_j \ket{\textbf{0}}$.
This allows us to relate the coefficient $c_j$ to 
the gate parameter $y_j$.
In practice, $c_j$ is in general a complex number while $y_j$ is a real parameter. Therefore, 
we assign two Pauli string operators to every single $j$ bit-string:
$P^{(r)}_j \ket{\textbf{0}}= i\ket{j}$ and $P^{(i)}_j \ket{\textbf{0}}= \ket{j}$.

If the RHS of Eq. (\ref{eq-MQITE-main-circuit}) is considered as a $2^n\times1$ vector $\sum_j \vert c_j\vert\, e^{i \theta_j} \ket{j}$, after $\chi$ times running and sampling the quantum processor, there are at most $\eta \le \chi$ non-zero entries (i.e., bit strings), with $\eta\ll 2^n$.
Therefore, it is computationally inexpensive to design a set of measurements and obtain both the \emph{amplitude} $\vert c_j\vert$ (already obtained from measurement in $Z$ basis) and the \emph{phase} $\theta_j$, for all the $\eta$ observed $j$ bit-strings. For more details see Appendix \ref{sec:MeasurementCircuit}.

From Eq. (\ref{eq-MQITE-main-circuit}) we can directly relate the real and imaginary parts of the coefficients $c_j$ to the gate parameters $y_j$'s. Explicitly:
\begin{eqnarray}
y^{(r)}_j = \delta\, c_j^{(r)}
\label{eq-y-real}
\end{eqnarray}
and 
\begin{eqnarray}
y^{(i)}_j =  -\delta\, c_j^{(i)},
\label{eq-y-imag}
\end{eqnarray}
where $y^{(r)}_j$ and $y^{(i)}_j$ are the parameters associated with the Pauli operators
$P^{(r)}_j$ and $P^{(i)}_j $, respectively.
The unitary operator $U_\delta$ is now determined up to $\delta$ order. 
Next, one replaces $U_{\tau}$ with 
$U_{\tau+\delta} = U_{\tau} U_\delta$, and proceed to the next step. 
The complete steps of the algorithm are shown in Fig. \ref{fig:MQITE}.

A different derivation of Eqs (\ref{eq-y-real}) and (\ref{eq-y-imag}) is provided in appendix \ref{sec:Derivation2} which 
takes the same approach as in \cite{Motta2019DeterminingEvolution}, which is done 
by introducing a cost function based on the difference between the unitary and non-unitary evolved states. In contrast, we show 
in Appendix \ref{sec:MeasurementCircuit}
that Eq. (\ref{eq-mqite-idea}) for the unitary evolution 
does not require solving equations on a classical computer.

In most real-world applications, the initial state is a product state. In quantum chemistry 
and for most fermionic problems the initial state 
is a Hartree-Fock state represented in terms of a second-quantized state (i.e., a Fock state). In spin problems such as Max-Cut, the initial state is one possible configuration of the vertices. It is straightforward to adjust the formalism 
introduced here to a case where the initial state is determined from $\ket{\textbf{0}}$
by some unitary 
transformation $U_m$; 
$
\ket{\tilde{\textbf{0}}} = 
U_m 
\ket{\textbf{0}}
$. This corresponds to initializing $U$ as $U=U_m$ instead of $U=I$ [see Fig. \ref{fig:MQITE}, line 1].

\subsection{Steps and specifications}
\label{sec-StepsAndSpecifications}
%
%
%

%%%%%%%%%%%%%%%%%%% Algorithm Flowchart
In this section we provide comments and explanation of the different 
steps of our algorithm, which is summarized in Fig. \ref{fig:MQITE}.
 The list of input parameters used in 
the algorithm and their descriptions are tabulated in 
Table \ref{tb-parameters}. 

%%%%%%%%%%%%%%%%%%% Algorithm Fig 1

% %%%%%%%%%%%%%%%%%%% Algorithm
% \begin{figure*}[h]
%     \centering
%     \includegraphics[width=\textwidth]{Figures/algorithmv2.eps}
%     \caption{The algorithm steps.}
%     \label{fig:MQITE}
% \end{figure*}
%%%%%%%%%%%%%%%%%%%%%%%%%%%%%%%%

\begin{figure}
    \centering
\begin{algorithm}[H]
\begin{algorithmic}[1]
\State $U \gets I $ 
\State $\ket{\psi_0} \gets  \ket{\textbf{0}} \equiv \ket{0_1 \cdots 0_n } $ 
\For{$t=1$ \textbf{to} $\frac{T}{n_H \delta}$ }: 
    \For{$k=1$ \textbf{to} $n_H$}: 
        \State $\delta_k \gets \delta \, w_k$ 
        \For{$r=1$ \textbf{to} $\chi \approx \mathcal{O}(10^{2\epsilon})$}: 
          \State \textbf{execute} $U^{\dagger} Q_k U \ket{\textbf{0}}$ \text{on a quantum computer}
            \State \textbf{record} \text{observed} $(\ket{j}, \vert c_j \vert )$ 
        \EndFor 
        \State \textbf{select} \text{dominant} 
        $\{ \ket{j} \}$  
        \Comment{$\eta = \vert \,\{ (\ket{j}, \vert c_j \vert ) \} \,\vert $}
        \State $U_{\delta_k} \gets I$ 
        \For{$\ket{j}$ \textbf{in} $\{ \ket{j} \}$}:
            \State \text{\textbf{measure} $c_j^{(r)}$ and $c_j^{(i)}$} 
            \Comment{Appendix B}
            \State $n_k \gets \sqrt{1 - 2\delta_k\ c_0 + \delta_k^2}$ 
            \State 
            $y_j^{(r)} \gets \delta_k c_j^{(r)}/n_k$; 
            $P_j^{(r)} \ket{\textbf{0}} = i\ket{j};$
            \State 
            $y_j^{(i)} \gets -\delta_k c_j^{(i)}/n_k$;
            $P_j^{(i)} \ket{\textbf{0}} = \ket{j};$
            \State $U_{\delta_k} \gets U_{\delta_k}\,
            [
            e^{
            i y_j^{(r)} \,
            P_j^{(r)} 
            }
             . 
            e^{
            i y_j^{(i)}\, 
            P_j^{(i)}
            }
            ]$ \Comment{Appendix A}
        \EndFor \vsp
        \State 
        $U \gets UU_{\delta_k}$
    \EndFor 
    \State \textbf{measure}\textbf{ observables:}  $E_{gs} = \langle0|U^\dagger H U|0\rangle$ \text{ etc.}
\EndFor
\end{algorithmic}
%%%%%%%%%%%%%%%%%%
% \caption{MQITE Algorithm}
\end{algorithm}
    \caption{MQITE Algorithm}
    \label{fig:MQITE}
\end{figure}

%%%%%%%%%%%%%%%%%%% Table Of Inputs
%%%%%%%%%%%%%%%%%%% Table Of Inputs
\begin{table}
{\setlength{\tabcolsep}{1em}
\begin{tabular}{|l |} 
 \hline
 \textbf{Algorithm Inputs and Parameters}:
  \\ 
  \hline \hline
  $\epsilon: $ Significant figures $\Leftrightarrow$ $10^{-\epsilon}$ precision.  
 \\
  $\eta: $ The maximum number of allowed components.  
 \\
   $H: $    $\{(Q_k, w_k)\}$; The Hamiltonian of the problem.
\\
  $n_H: $    $\vert \{(Q_k, w_k)\}\vert$; Number of terms in $H$.
\\
 $\delta: $    Imaginary-time increment
\\
 $T: $    Time of simulation 
\\
 $n: $ Number of qubits.
\\
\hline
 $\chi: $ Number of circuit executions per step; $\mathcal{O}(10^{2\epsilon})$
 \\
  $m_\tau: $  $\frac{T}{\delta}$  number of steps.
\\
\hline
\end{tabular}
}
\caption{List of the inputs and parameters used in the algorithm. $\chi$ and $m_{\tau}$ are implied by other inputs.}
\label{tb-parameters}
\end{table}

%%%% Hamiltonian
Throughout the algorithm the
Hamiltonian is assumed to be of the general form 
$H=\sum_{k=1}^{n_H} w_k\, Q_k$, where $Q_k$ 
is a Pauli string operator. 
Equivalently, the Hamiltonian is represented as the set 
$H \equiv \{(w_1, Q_1), \dots (w_{n_H}, Q_{n_H}) \}$. 

%%%% Initialization
The algorithm begins with initializing  
qubits and the quantum processor (line 1 and 2).
If a different initial state $\ket{\tilde{\textbf{0}}}$ is 
to be considered, a circuit instruction 
$U_m$ such that 
$
\ket{\tilde{\textbf{0}}} = 
U_m 
\ket{\textbf{0}}
$
must be provided; then, $U=I$ is replaced with $U=U_m$ on line 1.
%%%% The time period
The full time of evolution is assumed fixed and denoted by $T$. 

%%%% loop over n_H paulis
At every time step, 
there is a loop over 
$
(w_k,Q_k) \in \{(w_1,Q_1), 
\cdots 
(w_{n_H},Q_{n_H})\}
$; see line 4. 
For every pair $(w_k,Q_k)$
we aim at finding the equivalent circuit that replicates 
$e^{-\delta w_k Q_k} \ket{\psi_{\tau}}$. 
For this purpose, following Eq. (\ref{eq-MQITE-main-circuit}),
the circuit $U^\dagger\, Q\, U\ket{\textbf{0}}$ is executed $\chi$ 
times on a quantum computer; see lines 6-9. The outcome of measuring 
the qubits in the computational $Z$-basis is a distribution over 
bit strings $\ket{j}$. The circuit is executed
$\chi \approx \mathcal{O}(10^{2\epsilon})$ 
times. 

There are $\eta \le \chi$ number of components in 
$U^\dagger\, Q\, U\ket{\textbf{0}}$ that are dominant. In line 10 the quantum computer \emph{draws} the relevant 
components of $U^\dagger\, Q\, U\ket{\textbf{0}}$ from the  probability 
distribution  
${\rm Prob}(j) = \vert \bra{j} U^\dagger\, Q\, U\ket{\textbf{0}}\vert^2$ 
with precision 
$10^{-\epsilon}$.
We assume the dominant components to
correspond to the largest 
observed probabilities within the considered precision (or number of executions). In practice we set a cutoff for $\eta$ and 
therefore $\eta$ is an input to the algorithm. 
The $\eta$ bit strings with largest probabilities, or, equivalently, with largest amplitudes, are recorded for next steps; see lines 8 and 10. 

Lines 12-18 compute the real and imaginary parts of $c_j$ 
for every observed  $j$.
For this purpose, further quantum circuit executions
and instructions are needed (see line 13) for each $j$ in order to
compute all $\eta$ real and imaginary parts. A possible 
approach is introduced in Appendix \ref{sec:MeasurementCircuit}.

On line 17, unitary operators $e^{i y^{(r)}_j  P^{(r)}_j}$ 
and $e^{i y^{(i)}_j  P^{(i)}_j}$ 
are prepared in terms of one- and two-qubit gates,
and are appended to the current circuit $U$. Implementation of
$e^{i y^{(r/i)}_j  P^{(r/i)}_j}$ in terms of elementary gates, under assumption of fully connected hardware, is straightforward and provided in Appendix \ref{sec:Decomposition}.
Each unitary operator $e^{i y^{(r/i)}_j  P^{(r/i)}_j}$ 
adds a maximum of
$\mathcal{O}(2n)$ to the circuit depth.
In Fig. \ref{fig:MQITE}, the circuit parameters $y_j$ are computed  more generally, i.e., beyond the first order in $\delta$ [compare lines 15 and 16 with 
Eqs. (\ref{eq-y-real}) and (\ref{eq-y-imag})].

$P^{(r)}_j$ and $P^{(i)}_j$ are chosen straightforwardly as follows:  
Since the initial state is $\ket{\textbf{0}}$, consider a Pauli string
operator $P^{(i)}_j$ where all single-qubit operators 
are the identity $I$, except for qubits that correspond to an entry $1$ in the bit tring $\ket{j}$, 
where we replace $I$ with $X$. The 
$P^{(i)}_j$ satisfies
$P^{(i)}_j \ket{\textbf{0}}= \ket{j}$.
$P^{(r)}_j$ is constructed by replacing \emph{one} 
$X$ operator with $Y$.

Finally, on line 19, the quantum circuit instruction is updated. The algorithm proceeds to the next $(w,Q)$. 
Once all the terms in $H$ are accounted for,
$\ket{\psi_{\tau+\delta}} = U\ket{\textbf{0}}$ is an approximation to $e^{-\delta  H} \ket{\psi_{\tau}}$ that is obtained by unitary evolution from initial state $\ket{\textbf{0}}$. At this stage the relevant observables such as the total energy expectation value can be computed; see line 21.
\subsection{Errors and Computational Cost}
\label{sec-ComputationalCost}

An ideal ITE simulation implements
$
\ket{\psi_\tau} = e^{-\tau H} \ket{\psi_0}
/ \vert \vert e^{-\tau H} \ket{\psi_0} \vert \vert
$, which is typically done by dividing  $\tau$ into 
$m_\tau = \tau/\delta$ 
slices. In this ideal situation, only two steps are involved: 
(1) non-unitary propagation 
$\ket{\phi_{\tau+\delta}} = 
e^{-\delta H} \ket{\psi_\tau}$
 and (2) the normalization 
 $
\ket{\psi_{\tau+\delta}} 
= 
\ket{\phi_{\tau+\delta}}/
\vert \vert \ket{\phi_{\tau+\delta}} \vert \vert
$. The ITE algorithm consists of $m_\tau$ iterations over these two steps.

By expanding $\ket{\psi_{\tau}}$ in the exact eigenstate basis 
of $H$, it is straightforward to show that $\ket{\psi_{\tau}} $
converges to the exact ground state (or a superposition of degenerate 
groundstates) when $\tau^{-1} \ll \Delta_E$, where $\Delta_E$ is the minimum gap between ground state and excited state energies of the Hamiltonian. 
Therefore,
\begin{eqnarray}
m_\tau   &\gg& \frac{1}{\Delta_E \, \delta}
\label{eq-thermal-depth}
\end{eqnarray}
must be satisfied for a given (fixed) value of $\delta$. 
In an ideal scenario where $e^{-\delta H}$ can be implemented 
with $\mathcal{O}(1)$ resources, this means that a circuit depth of at least
$\mathcal{O}\left((\Delta_E \, \delta)^{-1}\right)$ is expected. 

In practice, $e^{-\delta H}$ must be discretized into pieces, 
a process known as trotterization:
\begin{eqnarray}
 e^{-\delta H}  &\approx& 
 e^{-\delta_1 Q_1} \cdots e^{-\delta_{n_H} Q_{n_H}},
 \label{eq-first-trotter}
\end{eqnarray}
where $\delta_k= \delta\, w_k$.
The above is not exact since 
$\delta_k \delta_{k^\prime} [Q_k , Q_{k^\prime}]\ne 0$ for most of  $Q_k$ and $Q_{k^\prime}$ in $H$, and therefore 
$
e^{-\delta_k Q_k 
\, 
-\delta_{k^\prime} Q_{k^\prime} } 
\neq 
e^{-\delta_k Q_k }
\, 
e^{
-\delta_{k^\prime} Q_{k^\prime} } 
$. The missing 
terms on the RHS of Eq. (\ref{eq-first-trotter}) result in \emph{fluctuations} around the ideal 
expected outcome 
$
e^{-\delta H} \ket{\psi_\tau}
 / \vert \vert e^{-\delta H} \ket{\psi_\tau} \vert \vert
$ at every time step. We refer to these fluctuation as  \emph{quantum fluctuations} and denote 
the associated energy scale -- 
the uncertainty in energy due to these fluctuation --   
by $\Delta_{\rm qfl}$.
For a fixed $\delta$, the energy uncertainty $\Delta_{\rm qfl}$
is not a function of number of steps $m_{\tau}$.
Therefore, even if Eq. (\ref{eq-thermal-depth}) is satisfied (for a fixed $\delta$) the quantum fluctuation can still induce an overlap with excited states.

There are two ways to reduce $\Delta_{\rm qfl}$. First, one can decrease $\delta$. Since
Eq. (\ref{eq-thermal-depth}) must be simultaneously satisfied, a larger 
$m_{\tau}$ is then required. Second, one can keep $\delta$ fixed but perform a higher-order trotterization, which adds more terms to the expansion of $e^{-\delta H}$ and 
thus increases the circuit depth as well. For either way, quantum fluctuations 
are reduced at the expense of increasing circuit depth. Notice that in this analysis the impact of the initial state on $\Delta_{\rm qfl}$ is not considered.

In order to  quantify the circuit depth 
at a fixed $\delta$ for a first-order trotterization, 
we need to access the required number of gates to implement 
$e^{-\delta\, Q}\, U \ket{\textbf{0}}$.
From Sec. \ref{sec-MainIdea}, 
we know that exactly $2 \eta$
unitary operators are required and must be
appended to the quantum circuit $U$, namely,  
$U_{\delta_k} = 
[
e^{i y_{j_1}^{(r)} \,P_{j_1}^{(r)} } \, . \, 
e^{i y_{j_1}^{(i)}\, P_{j_1}^{(i)} }
] \cdots 
[
e^{i y_{j_\eta}^{(r)} \,P_{j_\eta}^{(r)} } \, . \, 
e^{i y_{j_\eta}^{(i)}\, P_{j_\eta}^{(i)} }
]
$ (see Fig. \ref{fig:MQITE}).
Since each unitary operator $e^{i y_{j}^{(r/i)} \,P_{j}^{(r/i)} } $
needs at most $2n$ elementary gates (see Appendix \ref{sec:Decomposition}), $4 \eta n$ elementary gates are added to the current quantum 
circuit $U$ in order to
implement $e^{-\delta Q} U \ket{\textbf{0}}$. Furthermore, there are $n_H$ terms in $H$ and thus a total number of  
\begin{eqnarray}
n_{gates} = \mathcal{O}(4 \eta n n_H)
\label{eq-num-gates}
\end{eqnarray}
gates per imaginary time step are added to the quantum circuit.

While the above estimate is exact, the value $\eta$ is yet undefined. 
We conjecture that $\eta$ is a fraction of the full Hilbert space. This conjecture is based 
on the fact that the difference between the incrementally propagated state $\ket{\psi_{\tau+\delta}}$
and $\ket{\psi_{\tau}}$ is not significant. Here we elaborate on this statement further. We then justify a cutoff value for $\eta$ based on the argument provided. As a result, a practical upper bound on the number of gates is obtained.

The core task of the MQITE algorithm is to
approximate
$U_{\tau}^\dagger\, Q\, U_{\tau}\ket{\textbf{0}} $ at each imaginary time step, i.e., to determine the coefficients $c_j$ on the RHS of Eq. (\ref{eq-MQITE-main-circuit}). We expect to need only few of these coefficients. To understand the reason, consider a  \emph{local} interaction $Q$ (i.e., a Pauli string operator affecting only a small number of qubits). In this case, we expect
$Q \ket{\xi_0(\tau)} = \sum_{j=0}^{j=2^n} c_j \, \, \ket{\xi_{j}(\tau)} $
to result in a set of coefficients with amplitudes $\vert c_j \vert$ sharply distributed around a certain $j^\star$.
Therefore, at finite precision, rather than 
including all $2^n$ possible contributions to the sum, we need to consider only a small number of terms falling around the maximum of the distribution.

In fact, at both short and large imaginary time limits, we expect the coefficients $c_j$ to be concentrated around a single $j$ component. Consider the short-time regime $\tau=\delta$ and a Hamiltonian $H$ containing an $O(n)$ number of local interactions. Because
$U_{\delta}\approx I$, $U_{\delta}^\dagger\, H\, U_{\delta}\ket{\textbf{0}}$ results in a superposition of 
very few other bit strings within a very short Hamming distance of the initial bit string $j=0$. Thus the distribution of amplitudes $\vert c_j \vert$ is concentrated at around $j=0$ in the beginning of the imaginary time evolution.
At the opposite limit, at $\tau=\infty$, consider 
the state $U_{\infty}^\dagger\, H\, U_{\infty}\ket{\textbf{0}}$. This state is also expected to involve a superposition very narrowly centered at $\ket{\textbf{0}}$ since, ideally, 
$U_{\infty}\ket{\textbf{0}}$ is the ground state of $H$.

It is possible that along the path $\tau: 0 \to \infty $ the distribution of amplitudes $\vert c_j \vert$ broadens, deviating from what is observed at the two extreme limits. 
This behavior is captured by the dependence of the standard deviation of this distribution on the imaginary time $\tau$. But there are other characteristics of the distribution that reveal additional information about the dominant number of components, $\eta$. These characteristics also can be used as performance metrics. A specific indicator is the difference between the largest 
$\vert c^{\star}_j \vert$ and the next largest amplitude for a given $\tau$. 
We denote this quantity by $\Delta^\star$. In Sec. \ref{sec-sub-maxcut-TFIM}, Fig. \ref{fig:TFIM and MAXCUT}, it is shown that $\Delta^*$ clearly displays the transition from $\tau=0$ to $\tau=\infty$.
In essence, the change in the statistics of amplitudes $\vert c_j \vert$ captures both thermal and quantum fluctuations. 

%A physical picture emerges if one makes an analogy to phase transitions in quantum many-body systems. 
\begin{figure*}[t]
    \centering
    \includegraphics[width=\textwidth]{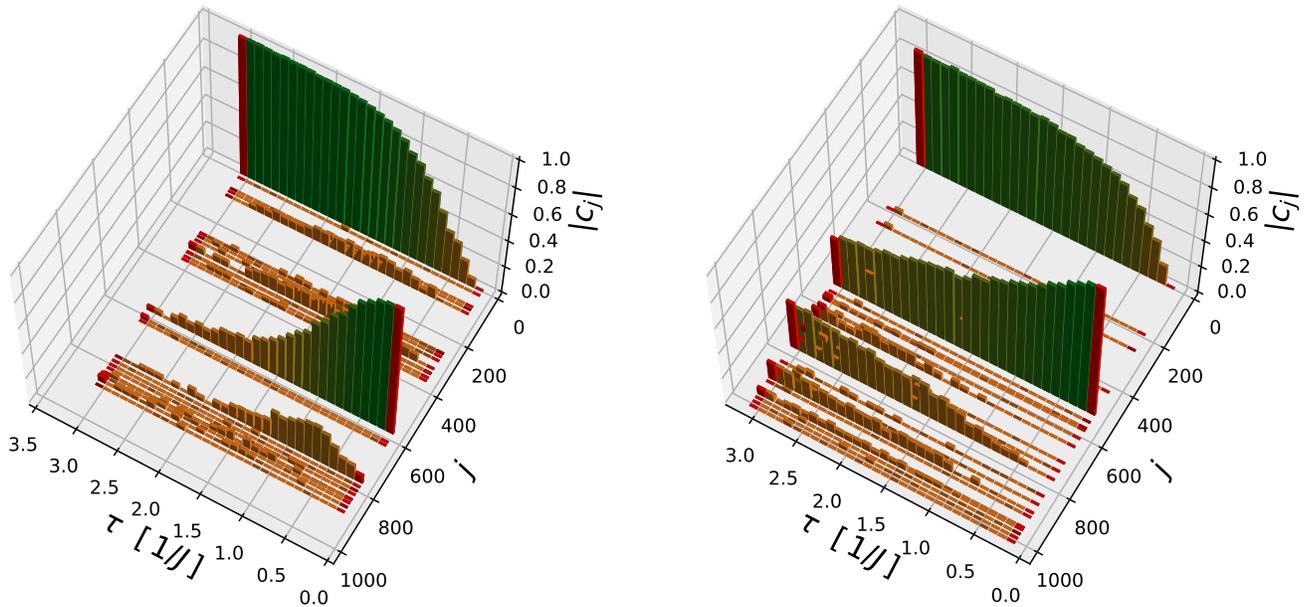}
    \caption{Evolution of the observed 
    $\{c_j (\tau) = 
    \bra{j}
    U_{-\tau} Q U_{\tau}
    \ket{\textbf{0}}
    \}$ distribution over the 
    $j$ indices for the last $Q$ term in $H$ and after $\chi=1000$ runs. (left) MaxCut problem with $n=10$ and (right) TFIM at the critical point for $n=10$ qubits. 
    At $\tau=\delta$ and $\tau=3$ the distribution is shown with different colors for clarity. At $\tau=\delta$, the distribution is sharply concentrated around a single $j$; a  Dirac-delta distribution. 
    As the evolution proceeds the distribution spreads out  and yet again approaches a Dirac-delta function around $j=0$. For Maxcut problem, all operators in $H$ commute, and thus a perfect Dirac-delta function is built on top of $j=0$ at $\tau=3$.
    Any discrepancy is either errors in MQITE simulation or persistent thermal fluctuation. In case of 
    Ising problem, quantum fluctuations slow down the expected process and 
    evolution toward a Dirac-delta function around $j=0$ is happening at a slower pace. More detail are provided in Sec. \ref{sec-sub-maxcut-TFIM} }
    \label{fig:evolution-of-cj}
\end{figure*}

In practice, the parameter $\eta$ is an input to the algorithm.
We make the assumption that $\eta$ is a function of $n$, such as $\eta = \alpha\, n^{d}$, 
where $d$ and $\alpha$ are constants. Thus $\eta$ plays the role of a cutoff and the number of gates can be upper bounded by 
\begin{eqnarray}
n_{gates} = \mathcal{O}(4 n^{d+1} n_H)
\label{eq-num-gates-bounded}
\end{eqnarray}
at any $\tau$.

Given $\eta$, to achieve a precision  $10^{-\epsilon}$, we estimate (see Appendix \ref{sec:MeasurementCircuit-Amplitude}) that $\chi \ge 10^{2\epsilon}\, \eta$
quantum circuit executions are needed, where $\epsilon$ is the desired
number of significant figures in the measurement of the amplitudes.

In Fig. \ref{fig:evolution-of-cj}, the evolution of the distribution of the amplitudes $\vert c_j (\tau) \vert$ is 
shown for the Max-Cut problem and the TFIM. These problems and the simulations are discussed in detail in Sec. \ref{sec-sub-maxcut-TFIM}. Here the distribution of amplitudes $\vert c_j (\tau) \vert$ 
is associated to the last $Q$ in $H$, that is, right before the loop on line 4 in Fig. \ref{fig:MQITE} is complete. At this point, the quantum state $U_{\tau}\ket{\textbf{0}}$ represents closely 
$e^{-\tau H}
\ket{\textbf{0}} 
/
\vert \vert 
e^{-\tau H}\ket{\textbf{0}}
\vert \vert$. 

At the boundary $\tau=\delta$, the distribution is sharply centered on a single $j$ index for both problems.
As the evolution  continues, the distribution spreads over up to $\eta$ components. In these simulations, we set $\eta=100=n^2$. 

At $\tau=3$, the maximum $\tau$ in these simulations, 
it is shown that the distribution $\{ \vert c_j (\tau) \vert \} $ narrows and peaks at $j=0$ again. For Max-Cut, where all the operators in $H$ commute, there distribution is more concentrated around a well-defined value. In the TFIM problem, due to non-commutation of the terms in $H$, there is a wider spreading. The latter is a consequence of the quantum fluctuations, where the uncertainty in energy allows some components of excited states to appear even at large $\tau$ (with fixed $\delta$).

\section{ Numerical Demonstration}
\label{sec-numerical}
In order to verify and demonstrate the MQITE algorithm, 
three different aspects are studied numerically.
All calculations are implemented using using 
the IBM Qiskit package \cite{Qiskit}. 
The first aspect relates to whether the algorithm presented in Appendix \ref{sec:MeasurementCircuit-Amplitude} can accurately compute
the real and imaginary parts of a given $c_j$ component. The second aspect is to test the algorithm on two toy problems that are 
frequently explored in literature, namely, Max-Cut and the TFIM, with a focus on 
the number of gates needed per time step. 
Finally, the last aspect concerns the application of the algorithm to a 
nuclear physics problem. In this problem, the Hamiltonian consists of 
Pauli strings that are not $k$-local, where $k$-local is defined by having any Pauli string operator $Q$ in $H$ to have at most $k$ operators from the set $\{X,Y, Z\}$, that is $Q$ acting on at most on $k$ qubits. The list of the Pauli string operators $\{Q\}$
and corresponding weights $\{w\}$ included in the Hamiltonian of this problem are tabulated in  Appendix \ref{sec:nuc-list-of-Qs}. The translations from a second-quantized Hamiltonian to Pauli string operators are performed by a Jordan-Wigner transformation \cite{jordan1993}.

%%%%%%%%%%%%%%%%%%%%%%%%%%%%%%%%%%%%%%%%%%%%%%%%%%%

\subsection{Validation of the Phase Estimation:}
\label{sec-sub-validation-phase}

In this numerical study, the goal is to illustrate that through the  
sub-algorithm presented in Appendix \ref{sec:MeasurementCircuit} one 
can estimate the phase $\theta$ of a component
such as $c_j= \vert c_j\vert e^{i\theta_j} = \bra{j} U^{\dagger} Q U \ket{\textbf{0}}$, or, equivalently, its real  and imaginary parts  $c^{(r)}_j$ and $c^{(i)}_j$, respectively.

A circuit represented by $U^{\dagger}\, Q\, U$ is prepared using Qiskit at every $\tau$
and for every $Q$ operator. The circuit is then executed $\chi=100$ times. 
For every execution, the qubits are measured
in the $Z$-computational basis and a bit string $j$ is recorded. After $\chi$ executions, 
a set of bit strings is identified; this is the set $\{j\}$. Let $\eta = \vert \{ j\} \vert$ be the number of observed components.
The maximum value of $\eta$ through the entire simulation is $\eta_{\rm max}=7$, which is much less than the $\chi=100$ shots used in the experiment. In the simulation, we obtain the 
value $\vert c_j \vert $ for every $j$ in the observed bit strings from the state vector representation of $U^{\dagger} Q U \ket{\textbf{0}}$, and round this value to  $10^{-\epsilon}$ precision. We note that to have $\vert c_j \vert$ determined with a precision of $10^{-3}$ requires $\chi \sim \mathcal{O}(10^6)$ circuit calls. 

The sets 
$\{\vert c_j \vert\}$ and  $\{j\}$ are used to 
obtain the real and imaginary parts (the phases) as explained in Appendix \ref{sec:MeasurementCircuit}:
For each $j$ component a separate circuit is prepared and executed $\chi$ times. However, since the number of runs necessary to reach the a desired precision (e.g., $\epsilon=2$) are too many given our computational resources, a smaller number of runs is used instead, and the amplitudes are read directly from the state vector and then rounded to precision ${\epsilon}$, thus mimicking the results that would be obtained following the measurement-based procedure.

\begin{figure*}[t]
    \centering
    \includegraphics[width=\textwidth]{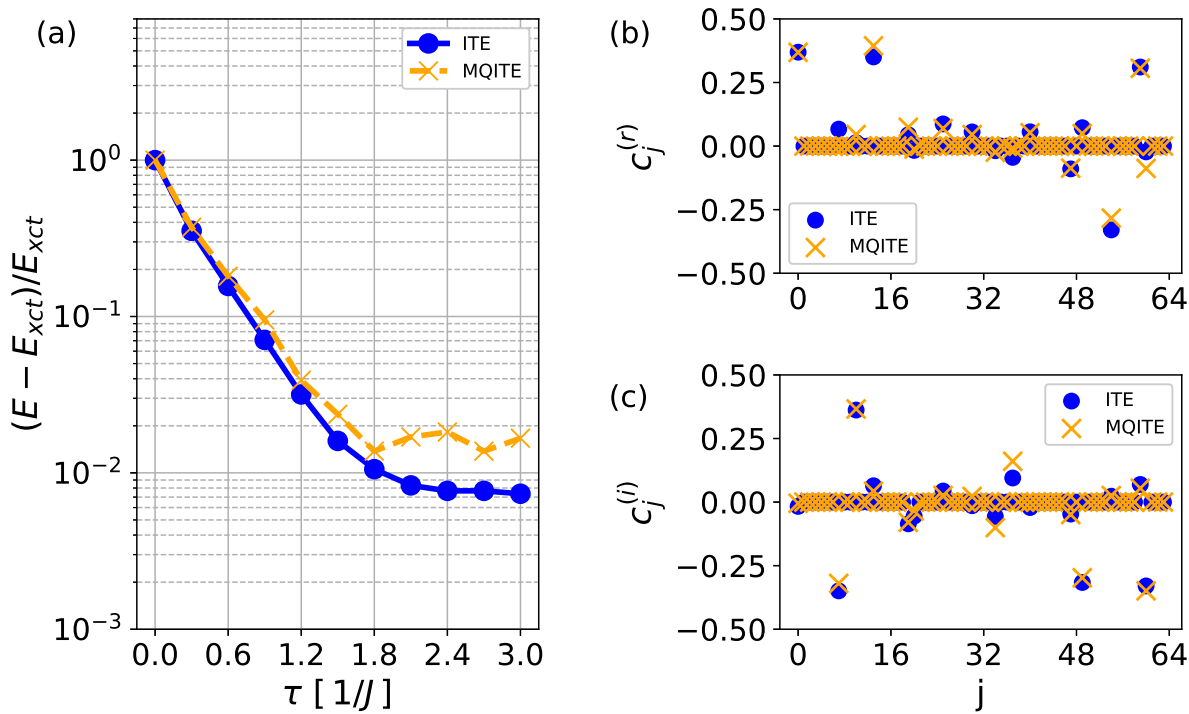} 
    \caption{
    (a) Relative difference betweem the computed ground state energy and the exact value for the ITE (blue circle) and MQITE (orange cross) methods as a function of the imaginary time. Real (b) and imaginary (c) parts of the components of the quantum state $\ket{\psi_{\tau}}$ at the final time evolution time step ($\tau=3/J$) for the ITE and MQITE methods. 
    }
    \label{fig:validation-of-phase}
\end{figure*}

The Hamiltonian is generated from 3-local interactions for a 6-qubit system. Thus every $Q$ acts on half of the qubits. The $Q$ operators are generated randomly from the set $\{ I, X, Y\}$. The Hamiltonian has the form
\begin{eqnarray}
H &=& J \left(0.961\,XXYIII + 0.853\,YIYIIY \right.
\nonumber \\
& & +\ \left. 0.137\,YIXYII + 0.980\,XIIIXY \right.
\nonumber \\
& & +\ \left. 0.712\,YIIIYX + 0.962\,XIYYII \right)
\end{eqnarray}
Throughout the simulation $\delta=0.3/J$ is used. Figure \ref{fig:validation-of-phase}a displays the difference between the computed ground state energy of the Hamiltonian and the exact value ($-3.118J$) as a function of the imaginary time for both ITE and MQITE. Both methods reach an accuracy of a few percent. Notice that the gap between the computed and the exact ground state energies tends to stabilize for both methods, an effect mainly due to the trotterization of the evolution. The gap is twice as large for MQITE, in part due to finite number of quantum measurement samples.

Figures \ref{fig:validation-of-phase}b,c 
show the real and imaginary parts of the components of the ITE
state vector 
$\ket{\psi_{\tau}^{\rm ITE}} = e^{-\tau H} \ket{\textbf{0}} / \vert \vert e^{-\tau H} \ket{\textbf{0}}\vert \vert$
and corresponding MQITE state vector
$\ket{\psi_{\tau}^{\rm MQITE}} = \sum_{j} \ket{j} \bra{j} U_{\tau} \ket{\textbf{0}}$
at the last imaginary-time evolution step of the simulation ($\tau=3/J$). Notice that only very few components are non zero. Real and imaginary parts obtained from the two methods match very closely.
Throughout the entire simulation the fidelity between the ITE and the MQITE state vectors stays above $0.998$.

\subsection{Max-Cut and Transverse-Field Ising Model}
\label{sec-sub-maxcut-TFIM}

In the following studies, two model problems are considered. The objective of the numerical experiments in this section is to show that the MQITE evolution matches with high fidelity the ITE evolution, ensuring that MQITE performs as well as the classical ITE method.

%%%%%%% Max-cut
\emph{Max-Cut}.-- The Max-Cut problem is to find the 
partitioning of a graph that maximizes the number of edges between the two partition. The problem maps onto a system of classical spins on a graph, where each spin interacts with $k$ other spins. An interaction is represented by an edge, and a spin by a vertex on a graph $G$ with a cost function given by
\begin{equation}
    H = \sum_{(tq)\in G} J_{tq} \, \, x_t\, x_q,
\end{equation}
where the weight $J_{tq}$ is the interaction strength between the spins at the vertexes $t$ and $q$ on the graph $G$. The classical spin $x$ takes the discrete values $\{-1, 1\}$. The ground state is a classical configuration
of the spin variables; the edges between spins with opposite values represent a
cut in the graph. The objective of the problem is to find the energy of the ground state. A quantum circuit solution to this problem was studied in Ref. \cite{Nishi2021ImplementationApproximation}, where the number 
of gates per time step was estimated to scale combinatorially with the number of bits $n$. To study this problem with a quantum algorithm, the $x$ variables are replaced by Pauli operators. In our study, each $x$ is replaced 
by a Pauli operator $X$. We randomly generated a graph with the property that each vertex is exactly linked to $k$ vertices; we set $k=3$. The weights $J_{tq}$ are chosen randomly and uniformly from the interval $\left[0, J\right]$.

%%%%%%% TISM 
\emph{TFIM}.-- We also study the well-known TFIM problem in one dimension. The Hamiltonian is given by
\begin{eqnarray}
H = -J \sum_{t =1}^{n-1} Z_t \, Z_{t+1} + h_{x} \sum_{t=1}^{n} X_t.
\end{eqnarray}
In our numerical studies, we set $h_x=J$, corresponding to the quantum critical point between paramagnetic and ferromagnetic ground states.

\begin{figure*}[t]
    \centering
    \includegraphics[width=\textwidth]{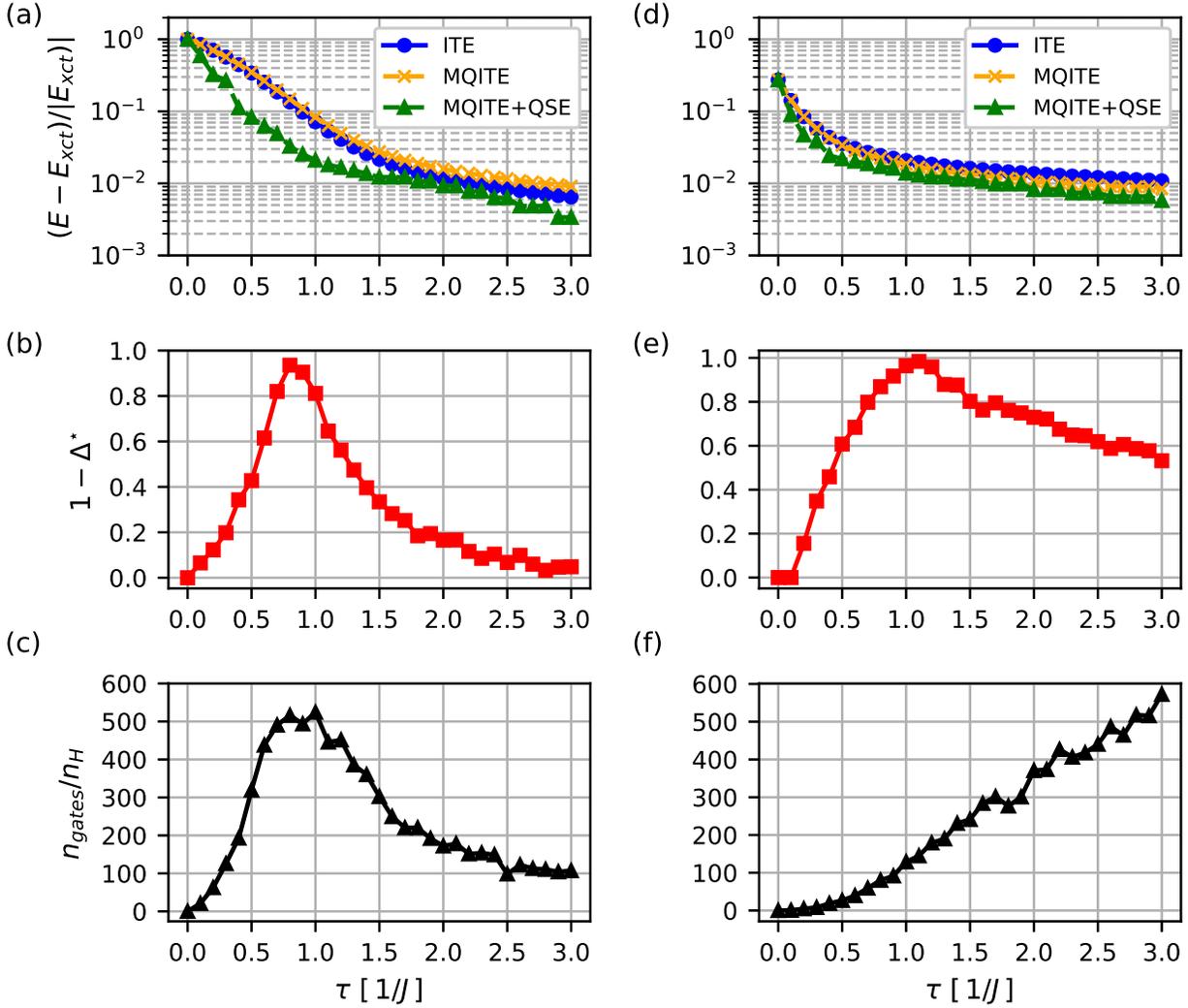} 
    \caption{
    The left panels correspond to the Max-Cut problem with $n=10$ and $k=3$; the right panels are for the one-dimensional TFIM at the critical point.  
    Panels (a) and (d) show the relative difference between the computed and the exact ground state energies for both ITE and MQITE. In addition, improved values after a QSE are shown (green triangle). Panels
    (b) and (e) show the dynamical change in $1-\Delta^{\star}$ for the two problems. Panels
    (c) and (f) show the number of added gates at every time step per number of terms in the Hamiltonian. All simulation are performed with a constant $\delta=0.1/J$ and $T=3/J$.
    }
    \label{fig:TFIM and MAXCUT}
\end{figure*}

%%%%%%%%%% 
%%%%%%%%%% inputs
%%%%%%%%%% 
\emph{Results}.-- For both Max-Cut and the TFIM, $n=10$ qubits were used. A 
circuit $ U^{\dagger} Q U$
was prepared using Qiskit and applied $\chi=1000$ times to the initial state $\ket{\textbf{0}}$. 
The observed set of $\{j\}$ bit strings corresponding to the first $\eta$ largest amplitudes $\vert c_j \vert$ was recorded for each value of $\tau$. As noted earlier, for practical reasons, the amplitudes were computed directly from the state vectors instead of  setting $\vert c_j \vert = \sqrt{n_j/\chi}$, where $n_j$ is the number of times the bit string $j$ is observed after $\chi$ executions of the circuit. The phases were also read off from the state vector and rounded to maintain $10^{-\epsilon}$ precision, with $\epsilon=2$, although the results are 
not sensitive significantly to the choice of $\epsilon$. $\eta$ was bounded by $n^2$ in these simulations. In the case of Max-Cut, the largest observed number of bit strings was $51$ for $\chi=1000$ and at $\tau=0.7/J$. In the case of the TFIM, the number was $38$ at $\tau=2.9/J$. The results of the numerical calculations are shown in Fig. \ref{fig:TFIM and MAXCUT}

%%%%%%%%%%
%%%%%%%%%% Slow convergence in 
%%%%%%%%%% TFIM
As Figs. \ref{fig:TFIM and MAXCUT}a,d show, the difference between the calculated and the exact ground state energies for the MQIT follows closely that for the ITE. We observe a slower convergence in the case of the TFIM for the considered values of $\delta$ and $T$. The slow convergence relates primarily to quantum fluctuation that emerges as a result of the trotterization. We have verified that the convergence improves upon either reducing $\delta$ or employing higher-order trotterization. Any further discrepancy between ITE and the MQITE is solely due to the algorithm itself. The most important source of such errors is the way the $\eta$ components are selected after $\chi$ executions. We choose the first $\eta$ components with the largest $\vert c_j \vert$ values, as justified in previous sections. Errors intrinsic to MQITE require further investigations.

%%%%%%%%%% Initializing in 
%%%%%%%%%% (1/sqrt(2)) * (|0>+|1>) 
%%%%%%%%%% if needed
The initial state of the qubits can improve the convergence of an ITE simulation, but may increase the number of terms in the set $\{j\}$ observed bit strings in MQITE. 
This increase in circuit depth was observed in the case of the TFIM when the initial stated was replaced by $(\ket{\textbf{0}} \pm  \ket{\textbf{1}}/\sqrt{2}$.

%%%%%%%%%% 
%%%%%%%%%% \Delta^\star
%%%%%%%%%% 
As argued in Sec. \ref{sec-ComputationalCost}, we expect the distribution 
of amplitudes $\vert c_j \vert $ to follow a particular dynamics. At the 
two extremes, $\tau=0$ and $\tau=\infty$, ideally, this distribution 
is narrowly concentrated at $j=0$. To illustrate this point, we 
define $\Delta^{\star}$ as the difference between the largest observed amplitude, $\vert c_j^{\star} \vert$, and the next largest amplitude in the distribution. Thus, we expect to obtain $\Delta^{\star}\approx 1$ at $\tau=0$ and $\tau=\infty$. Between these two extremes, the distribution is expected to broaden and 
thus $\Delta^{\star}<1$. The larger the number of bit string $j$ observed, the smaller the value of $\Delta^{\star}$. In Figs. \ref{fig:TFIM and MAXCUT}b,e, the difference $1-\Delta^{\star}$ is plotted versus imaginary the time. For Max-Cut this quantity is nearly zero at both ends of the evolution, with a relatively sharp peak developing at intermediate times. For the TFIM the same quantity evolves slower, with broad maximum and a wider tail at long times.

%%%%%%%%%% 
%%%%%%%%%% Number of gates
%%%%%%%%%% 
Figures \ref{fig:TFIM and MAXCUT}c,f show the number of gates added to the circuit at every time step divided by the number of terms in the Hamiltonian. The gate count is obtained from the Qiskit package and it agrees with the  estimate that for every $e^{iy_j P_j}$ operator used, there is approximately $2 n$ elementary gates from the set $\{R_x, R_y, R_z, CNOT\}$. Here it is assumed that CNOT gates between any two qubits are possible. 

%%%%%%%%%% 
%%%%%%%%%% QSE
%%%%%%%%%% 
In addition to MQITE, one can apply the quantum subspace expansion (QSE) introduced in Refs. \cite{McClean2017HybridStatesb,Motta2019DeterminingEvolution} (named QLanczos) to improve the accuracy of the ground state computation. This requires saving the 
circuit instruction of $U_{\tau}$ at all time steps, namely, saving the set $\{(y_j(\tau), P_j(\tau) )\}$ for all $\tau$.
% Using the non-orthogonal basis set $\{ U_{\tau} \ket{\textbf{0}} \} $, we applied the QLanczos algorithm introduced in .
This post-processing step requires solving a generalized eigenvalue problem using the correlation matrix 
$C_{\tau, \tau^\prime} 
= 
\bra{\textbf{0}}
U^{\dagger}_{\tau} 
\,
U_{\tau^\prime} 
\ket{\textbf{0}}$,
and an effective Hamiltonian. The effective Hamiltonian is defined within the Krylov subspace of $\{ U_{\tau} \ket{\textbf{0}} \} $
as
$H^{\rm eff}_{\tau, \tau^\prime} 
= 
\bra{\textbf{0}}
U^{\dagger}_{\tau} 
\,
H 
\,
U_{\tau^\prime} 
\ket{\textbf{0}}$,
where $H$ is the original Hamiltonian of the problem. Technical details can be found in the literature  \cite{tsukiyama2011medium, Cortes2022QuantumEstimation, Tkachenko2022} and are not provided in this paper. This quantum subspace expansion approach compliments ITE algorithm.

In Figs. \ref{fig:TFIM and MAXCUT}a,d, MQITE+QSE indicates the result of QSE-enhanced computations. Provided that enough memory to store all intermediate values for $\{(y_j(\tau), P_j(\tau) )\}$, there is a clear gain in post-processing with QSE.
% July 19: 
% Part of the plot is out of the frame
% the two subplots are too close
% 
\subsection{Application to Nuclear Physics}
\label{sec-sub-application-nuclear-physics}

\begin{figure*}[t]
    \centering
    \includegraphics[width=\textwidth]{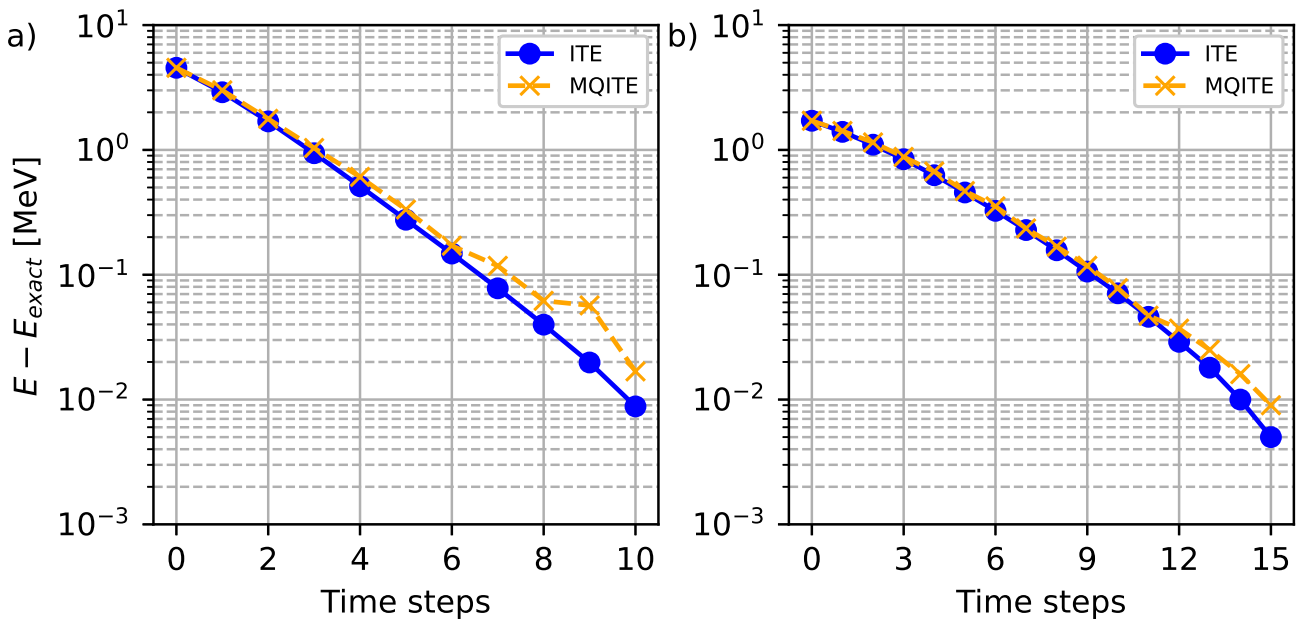} 
    \caption{
    Relative difference between computed and exact ground state energies as a function of imaginary time steps for two neutrons in the $0p_{3/2}$ and $0p_{1/2}$ orbitals interacting 
    via the Cohen-Kurath empirical interaction \cite{cohen1965effective}. 
    Panel (a) shows an initial configuration with the neutrons in the $0p_{3/2}$ 
    orbital, with $m=\pm 3/2$; this converges to the $J=0$ (zero angular momentum) ground state.
    Panel (b) shows an initial configuration with the neutrons in the $0p_{3/2}$ 
    orbital, with $m=+1/2, +3/2$, and hence total $M=2$; this converges 
    to the lowest $J=2$ state.
    Here $\delta = 0.05$ MeV$^{-1}$. 
    Although $n^2$ is the chosen cutoff for $\eta$, the largest observed $\eta$ is 
    $12$ and $14$ for (a) and (b), respectively. In this problem, $n=6$ and $\epsilon=3$.
    }
    \label{fig:nuclear}
\end{figure*}

A venue where this algorithm, and its lack of relying upon $k$-locality, is 
particularly appealing is in configuration-interaction calculations of the nuclear 
shell model~\cite{Caurier2005}. Here, the many-body basis states are in the occupation representation of Slater determinants, i.e., antisymmetric 
products of single-particle states of the form $ \hat{a}^\dagger_1 
\hat{a}^\dagger_2 \ldots \hat{a}^\dagger_N 
| 0 \rangle$, where the $\hat{a}^\dagger_i$
are fermion creation operators, and 
the second-quantized Hamiltonian has the general form
\begin{equation} 
\hat{\cal H} = \sum_{ij}T_{ij} \hat{a}^\dagger_i \hat{a}_j 
+ \sum_{i < j, k < l} V_{ij,kl}
\hat{a}^\dagger_i \hat{a}^\dagger_j 
\hat{a}_l \hat{a}_k.
\end{equation}
The matrix elements may encode important symmetries such as rotational 
invariance, but otherwise are not restricted, and hence the resulting matrix representation of the Hamiltonian in a many-body system is sparse but very nonlocal. In classical calculations, the low-lying eigenstates of the many-body 
Hamiltonian are efficiently found through 
the Lanczos algorithm or some of its variants~\cite{Caurier2005}.

Such configuration interaction calculations have many attractive qualities: 
the method (aside from the number of iterations to convergence) is not dependent upon the states or interaction having special properties; it is almost as easy to generate low-lying eigenstates as it is to generate the ground state; and it is 
not restricted to, for example, even or odd number of particles. The method is not, 
however, size-extensive, and the basis grows exponentially with the number of 
particles and/or the number of single-particle states. Because of this exponential growth, such number-occupation-based configuration-interaction calculations are obvious targets for quantum computation efforts. 

In Fig.~\ref{fig:nuclear}, we show results for two neutrons in the $0p$ shell, that is, with the $0p_{3/2}$ and $0p_{1/2}$ orbitals, which have a total of six single-particle states for protons and for neutrons, using the empirical Cohen-Kurath interaction~\cite{cohen1965effective}.
Because such nuclear shell-model Hamiltonians are rotationally invariant, 
the eigenstates have good total angular momentum $J$; for a given $J$, the eigenstates are degenerate in the $z$-component of angular momentum $M$.
Hence, in classical calculations, one often works in the so-called $M$-scheme basis,
where total $M$ is fixed.  We are able to apply this scheme here. Fig.~\ref{fig:nuclear}a has an initial configuration of two neutrons in the 
$0p_{3/2}$ orbital; one neutron has 
initial $m=+3/2$ and the other has $m=-3/2$, hence the total $M=0$. Under our simulation we recover the ground state which happens 
to have $J=0$. In Fig.~\ref{fig:nuclear}b, while the initial configuration still 
is two neutrons in the $0p_{3/2}$ orbital, the $m$-values are $+1/2,+3/2$, and 
hence the total $M=2$. Thus the system can only access states with total $J \geq 2$ (which in this case is the highest angular momentum available), and which is an excited state. Overall, there is a quite good agreement between ITE and MQITE for this problem, and a tendency for convergence to even more accurate ground state energy values for longer imaginary times. However, deviations do appear at longer times. As one may expect, noise can cause mixing with states of different $M$; we leave an investigation of this issue to future work.

\section{Conclusion and Discussion}
\label{sec-Conclusion}

In this section, we provide some comments about MQITE, its shortcomings and possible improvements, and its applicability to a wider range of problems than those studied in this paper.

\bigskip
%%%%%%%%%%%%%%%%%%%%%%%%%%%%%%%% Closing statement
%  PJ: This is what I had originally. I feel the meaning of this statement:
%
% "An ideal quantum computer or processor is expected to be working stand-alone. A
% computational overhead that is due to hybrid quantum-classical design is simply a step away from ideal quantum computers."
% 
%  is the opposite to the new statement (below) 
%
Ideally, a quantum computer should do as much possible the computations that are hard for a classical computer. A hybrid quantum-classical design where some hard computations are performed classically is a less desirable solution. 
 In the majority of variational quantum algorithms, the ansatz is a hypothesized parametric quantum circuit (PQC). A PQC is inevitably 
limited to a subspace of the full Hilbert space of the problem. Thus, for a 
generic problem where the solution is not in the image of the hypothesized  
ansatz, a classical optimizer or solver, no matter how precise it is, will never find the solution. 

In this paper a quantum algorithm for quantum state preparation is presented 
where the ansatz is not a hypothesized PQC to be optimized. In contrary, in our 
approach PQC is built gradually based on  direct measurements by the quantum 
computer, and without any classical optimizer or solver. Our algorithm is 
specifically demonstrated to implement ITE on a quantum computer where the target quantum state is the ground state. 

%%%%%%%%
%%%%%%%% no parameter - parameter correlation
%%%%%%%%
In general, variational quantum algorithms require optimization. Optimization of a PQC can be difficult, as the gate parameters are not in general independent. Our algorithm builds the circuit layer by layer in a way that parameters in each layer are uncorrelated by construction -- a consequence of working with an orthogonal basis set $\{\ket{\xi} \}$ where each basis state has an associated gate parameter.

%%%%%%%%%%%%%%%%%%%%%%%%%%%%%%%% Applications
Similarly to QITE, MQITE can be used as basis for a subspace expansion to obtain excited states.
%%%%%%%%
%%%%%%%% Nuclear; k-local
%%%%%%%%
One of the advantages of complementing the MQITE with the Lanczos algorithm 
is the ability to rapidly converge toward low-lying excited states. This is particular important in nuclear physics, where  low-lying spectra and transitions reveal much about the structure of nuclei. Here, we illustrated this strategy for the Max-Cut and the TFIM; we leave its application to nuclear physics problems to a future work.

%%%%%%%%
%%%%%%%% Applications ML 
%%%%%%%%
Other potential application of the MQITE algorithm is in machine learning (ML). Generally speaking, a typical task in ML is classification and prediction on incoming data, when the ML model is trained on some historical data. The algorithm presented here is anticipated to facilitate the preparation of input data as part of an quantum ML program.

%%%%%%%%
%%%%%%%% Applications Fusion 
%%%%%%%%
Another area of applicability is in fusion energy science where the objective are mostly simulation of flows of Navier-Stokes fluids and equations \cite{FrankG2020,  LloydNDE2020}. Recently, a quantum algorithm was presented to simulates a single time step of a Navier-Stokes fluid using a quantum simulator \cite{METHODLJUBOMIR2022}. In that simulation, at every time step the circuit was initialized to the quantum state obtained in the previous step. 
In our approach, working in a different basis 
($\{ \ket{\xi_j}\} $) rather than in the computational basis allows us to control the increase in complexity from one step to another. We thus expect MQITE to be a useful tool for simulations in the fusion energy area of research.

%%%%%%%%%%%%%%%%%%%%%%%% improving / not considered
%%%%%%%%
%%%%%%%% improving the quality by better sampling- ML
%%%%%%%%
In the current version of the algorithm, at every time step, the most dominant components are deemed to be the ones that have the
largest amplitudes. This assumption may be intuitive, but is not necessarily and generally correct in every situation. An ML-assisting software could possibly infer a better image of the distribution of amplitudes $\{ c_j \}$ that is obtained from measuring $U^{\dagger} \, Q\, U \ket{\textbf{0}}$, and thus improve the accuracy of MQITE.

%%%%%%%%
%%%%%%%% gate errors
%%%%%%%%
In our studies, the effect of hardware errors were not considered. However, we specifically take into consideration the precision required in the quantum measurements to achieve a certain precision in the computation. The constraints on the measurements translate directly into gate parameters. Hence it is implicitly assumed that any one- and two-qubit quantum gate can be performed up to the desired precision.

%%%%%%%%
%%%%%%%% effect of \delta and T; no nisq
%%%%%%%%
Intrinsic to the classical ITE approach and to trotterization, an increase in the number of imaginary time steps often results in an increase in circuit depth. The number of gates per time step in our algorithm is estimated to be polynomial in the system size (number of qubits) for a fixed, reasonable precision. Nevertheless, even if such a favorable scaling, since relatively large circuit depths are expected even for low precision, the MQITE algorithm is not suitable for current NISQ systems.

\section{Acknowledgements}
This material is based upon work supported by the U.S. Department of Energy, Office of Science, Office of High Energy Physics, under Award Number(s) DE-SC0019465; Office of Basic Energy Sciences under Award Number(s) DE-SC0019275; Office of Fusion Energy Sciences, under Award Number(s) DE-SC0020249; and the auspices of the National Nuclear Security Administration of the U.S. Department of Energy at Los Alamos National Laboratory, under Award Number(s) 89233218CNA000001 as well as partial support by the Advanced Simulation and Computing (ASC) Program. The authors thank Peter Smucz for graphical support, Mark Kostuk, Stefan Bringuier, and Matthew Cha at GA, and Yuri Alexeev and Dmitry Fedorov at Argonne National Laboratory (ANL) for discussions.

% The work of CWJ was supported by the U.S. Department of Energy, Office of Science, High Energy Physics, under Award DE-SC0019465. The work of IS was carried out under the auspices of the National Nuclear Security Administration of the U.S. Department of Energy at Los Alamos National Laboratory under Contract No. 89233218CNA000001. IS gratefully acknowledges  partial support by the Advanced Simulation and Computing (ASC) Program. ERM was partially supported by the the U.S. Department of Energy, Office of Science, Basic Energy Sciences, under Award DE-SC0019275.
% % PJ
% PJ's material is based upon work supported by the U.S. Department of Energy, Office of Science, Office of Fusion Energy Sciences, using the DIII-D National Fusion Facility, a DOE Office of Science user facility, under Award Number(s) DE-SC0020249.
% %
% The authors thank Peter Smucz for graphical support. PJ thanks Mark Kostuk, Stefan Bringuier, and Matthew Cha at General Atomics, and Yuri Alexeev and Dmitry Fedorov at Argonne National Laboratory for discussions.
%% in preparation keep this uncommented and run
\bibliography{Bibtex/references_pejman,Bibtex/johnsonmaster.bib}

\clearpage

\appendix
%%%%%%%%%%%%%%%%%%%%%%%%%%%%%%%%%%%%%%%%%%%%%%%%%%%%%
%%%%%%%%%%%%%%%%%%%%%%%%%%%%%%%%%%%%%%%%%%%%%%%%%%%%%
%%%%%%%%%%%%%%%%%%%%%%%%%%%%%%%%%%%%%%%%%%%%%%%%%%%%%
%%%%%%%%%
%%%%%%%%%
%%%%%%%%%.  Decomposition Oracle
%%%%%%%%%
%%%%%%%%%
%%%%%%%%%%%%%%%%%%%%%%%%%%%%%%%%%%%%%%%%%%%%%%%%%%%%%
%%%%%%%%%%%%%%%%%%%%%%%%%%%%%%%%%%%%%%%%%%%%%%%%%%%%%
%%%%%%%%%%%%%%%%%%%%%%%%%%%%%%%%%%%%%%%%%%%%%%%%%%%%%

\section{ Decomposition Oracle}
\label{sec:Decomposition}
Consider the following unitary that acts on three qubits:
%%%%%
%%%%% A1
%%%%%
\begin{eqnarray}
U^{(3)}\,(\alpha) = \cos{\alpha} \,\,\,I + i\sin{\alpha}\,\,\, X_1 X_2 X_3.
\nonumber \\
\label{eq-3qubit}
\end{eqnarray}
For any  unitary 
operation $R$  we have
%%%%%
%%%%% A2
%%%%%
\begin{eqnarray}
 U^{(3)}(\alpha) =  
 R^{\dagger}
 \left[
 R  \,U^{(3)}(\alpha) R^{\dagger}
 \right]
 R.
\end{eqnarray}
Now, consider $R$ such that the term within the brackets reduces to
%%%%%
%%%%% A3
%%%%%
\begin{eqnarray}
R  U^{(3)}\,(\alpha)  
R^{\dagger} = 
I
\otimes 
U^{(2)}(\alpha),
\label{eq-ex-3}
\end{eqnarray}
where $U^{(2)}$ acts on two qubits.
Then, the number of operating gates is reduced from three qubits to two qubits.
If the operation $R$ is chosen such that it only affects one or two qubits, then the full circut consists of only of one- and two-qubit gates.

\begin{figure*}[t]
    \centering
\begin{quantikz}
\\
\lstick{$\ket{i}$} 
& \gate{u_{\frac{\pi}{2}, \frac{-\pi}{4}, 0} } 
& \ctrl{1} 
& \gate{u_{\frac{\pi}{2},
\frac{\pi}{2}, 
\frac{-\pi}{4}} } & \qw \\
%%%%
\lstick{$\ket{i+1}$} 
& \gate{u_{\frac{\pi}{2}, -\pi, 0} } 
& \targ{} 
& \gate{u_{
\frac{\pi}{2}, 
\frac{\pi}{2},
0} } & \qw 
\\
\end{quantikz}
\caption{Decomposition of the $R$ gate in terms of elementary gates.}
\label{fig:R-gate-decomposition}
\end{figure*}
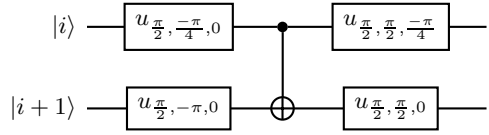

Following this result, 
suppose an application of 
a unitary $R$ allows us to
reduce $U^{(n)}$ to $U^{(n-1)}$. 
By recursion, we can construct an equivalent circuit $U^{(n)}$ that consists of 
$\mathcal{O}(n)$ single-qubit and two-qubit gates only. Namely,
%%%%%
%%%%% A4
%%%%%
\begin{eqnarray}
U^{(n)}(\alpha) = R_{1,2}^\dagger 
\, \cdots \, 
U^{(2)}\,(\alpha) 
\cdots \, R_{1,2},
\nonumber \\
\label{eq-generic}
\end{eqnarray}
where $R$ is a gate that acts on 
two qubits.
A candidate for $R$ is
%%%%%
%%%%% A5
%%%%%
\begin{eqnarray}
 R_{i, i+1} =  
 \frac{1}{2}
 (I + iZ_{i+1} )
 (I - i X_{i}Z_{i+1}).
 \nonumber \\
 \label{eq-R}
\end{eqnarray}
The explicitly decomposition of $R$ in terms of single-qubit rotation $u_{\theta_x, \theta_y, \theta_z}$
and C-NOT gate is shown 
in Fig. \ref{fig:R-gate-decomposition}.

%%%%%
%%%%% Fig 6
%%%%%
%%%%%%%%%%%%%%%%%%%%%%%%%%%%%%%%
%\begin{center}
%\end{center}
%%%%%%%%%%%%%%%%%%%%%%%%%%%%%%%%

\begin{figure*}
%%%%%%%%%%%%
\begin{quantikz}
%%%%%%%%%%%
%%%%%  q 1
\lstick{$1$} 
& \gate{u_1} 
& \gate[2,disable auto height]{\begin{array}{c}
 R
\end{array}} 
& \qw 
& \qw
& \qw 
& \gate[2,disable auto height]
{
\begin{array}{c}
R^\dagger
\end{array}
} 
& \gate{u_1^\dagger} 
\\
%%%%%  q 2
\lstick{$2$} 
& \gate{u_2}
&
& \gate[2,disable auto height]
{
\begin{array}{c}
R
\end{array}
} 
& \qw 
& \gate[2,disable auto height]
{
\begin{array}{c}
R^\dagger
\end{array}
}
& 
& \gate{u_2^\dagger} 
\\
%%%%%  q 3
\lstick{$3$} 
& \gate{u_3}
& \qw 
& 
& \qw 
& 
& \qw 
& \gate{u_3^\dagger}  
& 
\\
\vdots 
&  
&  
& 
& \vdots  
&
&
\\
%%%%%  q n-1
\lstick{$n-1$} 
& \gate{u_{n-1}}
& \qw  
& \qw 
& \gate[2,disable auto height]
{
\begin{array}{c}
U^{(2)}(\alpha)
\end{array}
}
& \qw
& \qw
& \gate{u^\dagger_{n-1}}
\\
%%%%%  n
\lstick{$n$} 
&\gate{u_{n}}
& \qw  
& \qw  
& 
& \qw
& \qw
& \gate{u^\dagger_{n}}
%%%%%%%%%%%
\end{quantikz}
%%%%%%%%%%
\caption{Implementation of a multi-qubit gate [Eq. (\ref{eq-multigate})]. The circuit begins with single-qubit change of basis, using the gate $u$, followed by a two-qubit gate $R$.
Once the parametric two-qubit gate $U^{(2)}$
is applied, the inverse of $R$ is applied. Finally, the one-qubit basis is returned to original one.
This is a special case where none of the $n$ operators $O$ in Eq. (\ref{eq-multigate}) is an identity $I$. Otherwise, some $R$ and $u$ gates are absent, and the $R$ gate may act on distant qubits. In this case, the circuit depth decreases further to $\mathcal{O}(\le 2n) $. }
\label{fig-decomp-multi-gate}
\end{figure*}
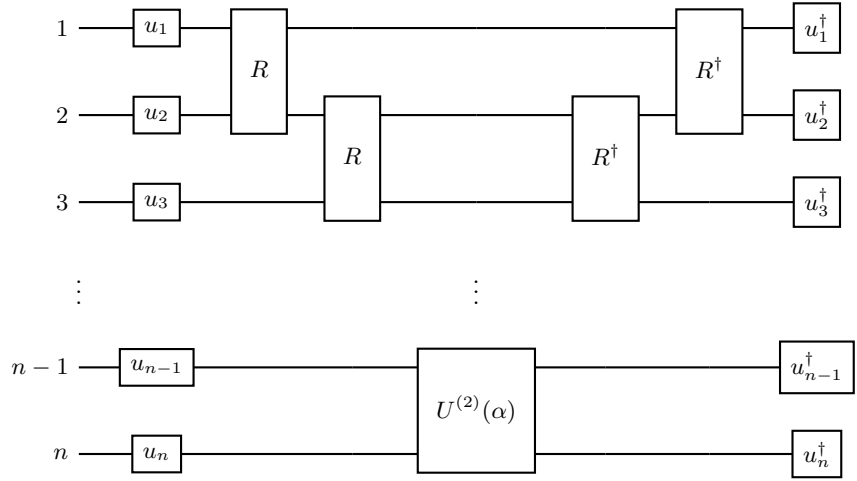

The breakdown 
of the full decomposition of
a unitary,
%%%%%
%%%%% A6
%%%%%
\begin{eqnarray}
U^{(n)}\,(\alpha) = 
\cos{\alpha} \,\,I 
+
i\sin{\alpha}\,\, O_1 \cdots O_n,
\nonumber \\
\label{eq-multigate}
\end{eqnarray}
where $O$ is any of the 
three Pauli operators, is  
depicted in Fig. \ref{fig-decomp-multi-gate} as a quantum circuit. A set of (non-parametric)
single-qubit gates ($u$ gates) are required
in order to turn $O$ into $X$ and vice versa at the beginning and final stages of the circuit.
The circuit depth is of order of $O(2n)$. 
%%%%%
%%%%% fig 7
%%%%%
%%%%%%%%%%%%%%%%%%%%%%%%%%%%%%%%
%\begin{center}
%\end{center}
%%%%%%%%%%%%%%%%%%%%%%%%%%%%%%%%

% Notice that a fully connected hardware achitechure is assumed. In particular for the scenario where some intermediate $O_i$ pauli's in Eq. (\ref{eq-ex-3}) are identity,
%  $\mathcal{R}$ acts on physically separated 
%  qubits. For a hardware with limited connectivity qubit swaps are to be applied, which increases the circuit depth. 

\section{ Measuring the Amplitude and the Phase}
\label{sec:MeasurementCircuit}
Consider $U$ 
to be a known circuit,  $Q$ a Pauli string operator, and the quantum state $\ket{\psi} = U^\dagger\, Q\, U \ket{0}$.
We conjecture that this state has a limited number of components in the computational bases. That is,
\begin{eqnarray}
\ket{\psi} &=& \sum_{j} c_j \ket{j}
\nonumber \\
&=& c_{j_1} \ket{j_1} +\cdots c_{j_{\eta}} \ket{j_{\eta}}
\label{eq-psi_UQU}
\end{eqnarray}
with $\eta \ll 2^n$, where $n$ is the number of qubits and $j$ in $\ket{j}$ 
can be any bit string between $0$
($\ket{0_1, \cdots, 0_n}$) and $2^{n-1}$ ($\ket{1_1, \cdots, 1_n}$ ). The component $c_j=\vert c_j\vert e^{i\theta_j}$ 
is a complex number and is specified by the amplitude $\vert c_j\vert$ and the phase $\theta_j$, or, equivalently, by the real and imaginary parts 
$c^{(r)}_j=\vert c_j\vert \cos{\theta_j}$ and 
$c^{(i)}_j=\vert c_j\vert \sin{\theta_j}$, respectively. In this Appendix, we show a way to obtain these two quantities using only a quantum computer, i.e, without classical optimization, up to a precision $10^{-\epsilon}$.

%%%%%%%%%%% Amplitudes
\subsection{Amplitude $|c_j|$}
\label{sec:MeasurementCircuit-Amplitude}
Formally, the procedure is to run the circuit $U^\dagger\, Q\, U$ on $\ket{0}$ a $\chi$ number of times, , and each time measure the qubits in the
computational basis. Then, $\vert c_j\vert = \sqrt{n_j/\chi}$, where $n_j$ is the number of times the string $j$ is observed. The precision in the determination of $\vert c_j\vert $ depends on $\chi$ and $\eta$: employing the central-limit theorem, it is possible to show that 
\begin{eqnarray}
\epsilon \approx \frac{1}{2} \log_{10}
 \left( \chi / \eta  \right).
\label{eq-precision}
\end{eqnarray}
When the circuit $U^\dagger\, Q\, U$ is executed 
$\chi$ times, at every shot the circuit collapses onto one of the $2^n$ possible bit strings. 
% $\chi$ is
% commonly referred to as \emph{shot} number (or shots), or the number of calls to the quantum processor. 
We conjecture that the number of distinct bit strings observed in each shot is $\eta \ll 2^n$, and possibly $\eta \sim {\cal O}(n^d)$. Therefore, to achieve a $10^{\epsilon}$ precision, we need $\chi \ge \eta\, 10^{2\epsilon}=\mathcal{O}(10^{2\epsilon})$ shots.

\subsection{Phase $\theta_j$}
\label{sec:MeasurementCircuit-Phase}
In order to obtain the phase without accessing the state vector, we propose the following 
sub-algorithm. This sub-algorithm is used and tested in Sec. \ref{sec-sub-validation-phase}.

We define the unitary $T_{ts}=e^{i\frac{\pi}{4} P_{ts}}$, where $P_{ts}$ is a Pauli string operator comprised of $X$ and $I$ such that $P_{ts}\ket{t}=\ket{s}$, where $s$ and $t$ are bit strings. $T_{ts}$ can be implemented  
using the decomposition scheme introduced in Appendix  \ref{sec:Decomposition}.
Using $T_{ts}$ and Eq. (\ref{eq-psi_UQU}), consider
\begin{eqnarray}
\ket{\psi^{\prime}} &=&T_{j_1 j_2} \left[U^\dagger Q U \ket{0} \right]
\nonumber \\
&=&(\frac{1}{\sqrt{2}} + i \frac{1}{\sqrt{2}} P_{j_1 j_2}) 
\, \left[\sum_{j} c_j \ket{j} \right]
\nonumber \\
&=& \frac{1}{\sqrt{2}}(c_{j_1} + ic_{j_2} ) \ket{j_1} +\cdots.
\label{eq-circ-theta}
\end{eqnarray}
Upon repetitive execution and measurement of the qubits in the $Z$-computational basis of single qubits, one can approximately obtain
 $m_{j_1}=\vert\bra{j_1}\ket{\psi^{\prime}}\vert^2$ approximately, that is, this quantity can be obtained with a precision specified by the number of shots. 
Since $m_{j_1} = (1/2) \vert c_{j_1} + i c_{j_2} \vert^2$, where 
the amplitudes $\vert c_{j_1}\vert$ and 
$\vert c_{j_2}\vert$ are already known from the execution of $\ket{\psi} = U^\dagger Q U \ket{0} $, by measuring $m_{j_1}$ the relative 
phase $\theta_{j_1} - \theta_{j_2}$ is obtained,
\begin{eqnarray}
\sin{(\theta_{j_1} - \theta_{j_2})} = 
\frac{
2m_j 
-\vert c_{{j_1}} \vert^2 
-\vert c_{j_2} \vert^2 
}
{2 \vert c_{{j_1}} \vert 
\vert c_{j_2} \vert }.
\label{eq-relative-theta}
\end{eqnarray}
Furthermore, assuming $\theta_{j_1}=0$, $\theta_{j_2}$ is obtained from the above equation. 

In order to obtain all  $\{\theta_{j_2} , \cdots \theta_{j_{\eta}} \}$, the the above experiment is repeated $\eta-1$ times, that is, $\eta-1$ different quantum circuits are employed, each time with a different 
$T_{j_1 j}$ operator,
$j \in \{ j_2, \cdots, j_{\eta} \}$.

While this approach is in principle sufficient for finding the phases, in practice we employed a modified approach. The objective for this modification is 
to avoid reliance on the relative phase between two phases $\theta_{j_1}$ and $\theta_{j_2}$, for the reason that the original approach may result in  different global phases from one iteration of the algorithm to the next; that is, the global phase of $U^\dagger Q U \ket{0}$ in the inner loop on line 4 of Fig. \ref{fig:MQITE} varies. 
%%%%%%
%%%%%% Why it should not matter
%%%%%%
In theory this should not be an issue, but we have numerical evidence  that this can be a source of error. Further investigation is needed to clarify this issue.

The modified approach is to assign the zero phase to a component that is not present in the set of $\{c_j\}$, in Eq. (\ref{eq-psi_UQU}). In other words, a $j_1$ is chosen such that $c_{j_1}=0$ (within the assumed precision),
and for this bit string  $\theta_{j_1}=0$ is assumed. However, a glance at Eq. (\ref{eq-relative-theta}) reveals that this results in a divergence, unless 
the limit of $c_{j_1} \rightarrow 0^{+}$ is taken. W then have
\begin{eqnarray}
\lim_{c_{j_1} \to 0^{+}} 
\frac{
2m_j 
-\vert c_{{j_1}} \vert^2 
-\vert c_{j_2} \vert^2 
}
{2 \vert c_{{j_1}} \vert 
\vert c_{j_2} \vert } &=&  
\frac{c_{{j_2}}^{(i)}  }{\vert c_{{j_2}} \vert  },
\label{eq-relative-theta-theory}
\end{eqnarray}
which is equivalent to the $\sin{(\theta_{j_2})}$. The limit  $c_{j_1} \rightarrow 0^{+}$
can be implemented with a quantum circuit that utilizes an ancillary qubit, as we shown below. 

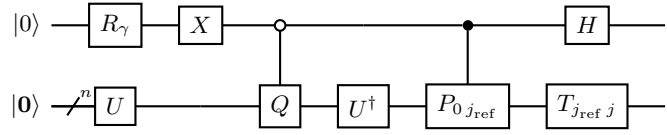
\begin{figure*}[t]
    \centering
    \begin{quantikz}
        \\
        \lstick{$\ket{0}$} 
        & \gate{R_{\gamma} } 
        & \gate{X} 
        & \octrl{1} 
        & \qw
        & \ctrl{1} 
        & \gate{H}  
        & \qw \\
        %%%%
        \lstick{$\ket{\textbf{0}}$} 
        & \gate{U} \qwbundle{n}
        & \qw
        & \gate{Q} 
        & \gate{U^\dagger} 
        & \gate{P_{0 \, j_{\rm ref}}}
        & \gate{T_{j_{\rm ref}\, j}} 
        & \qw
        \\
        \end{quantikz}
    \caption{Quantum circuit to measure the imaginary part of $c_j$ (line 12 in Fig. \ref{fig:MQITE}). For each 
    bit string $j$ observed in the 
    execution of $U^\dagger Q U \ket{\textbf{0}}$, a measurement circuit as the above is prepared and used to obtain the associated $c_j^{(i)}$ part. 
    Here, $P_{0 \, j_{\rm ref} }$ is a Pauli string operator with only $I$ and $X$ Pauli operators such that 
    $P_{0 \, j_{\rm ref} }\ket{0}=
    \ket{j_{\rm ref}}$.
    }
    \label{fig:mesurment-phase-imag}
\end{figure*}

\emph{Imaginary Part}.--
Consider the circuit in Fig. \ref{fig:mesurment-phase-imag}. Here $R_{\gamma}$
is a rotation on the ancillary qubit, with $\gamma$ being a small positive value. In practice, $\gamma=10^{-\epsilon}$ is used and
$R_{\gamma} $
is defined such that
$
R_{\gamma} 
\ket{0} = 
\cos{(\gamma)}
\ket{0} + \sin{(\gamma)} \ket{1}.
$
%%%%%
%%%%% fig 8
%%%%%
%%%%%%%%%%%%%%%%%%%%%%%%%%%%%%%%
%\begin{center}
%\end{center}
%%%%%%%%%%%%%%%%%%%%%%%%%%%%%%%%
The quantum circuit in Fig. (\ref{fig:mesurment-phase-imag}) 
performs the following unitary transformation:
%%
%%%%%
%%%%%%%
\begin{eqnarray}
\ket{0_a} \ket{\textbf{0}} &\rightarrow&
\ket{0_a} \ket{j_{\rm ref}} 
\frac{1}{2}
\left[ 
i \cos{(\gamma)} c_{j} 
+ 
\sin{(\gamma)} 
\right] 
+ 
\cdots
\nonumber \\
&\approx&
\ket{0_a}  \ket{j_{\rm ref}}
(\frac{i}{2})
\left[ 
c_{j} - i\gamma
\right]
+ \cdots
\nonumber \\
&=& \ket{\Psi^\prime}
.
\label{eq-circ-theta-ancila}
\end{eqnarray}
%%%%%%%%
%%%%%%
%
Here $j_{\rm ref}$, is a bit string that is not in the set $\{j_1, \cdots j_{\eta} \}$ associated to $ U^\dagger Q U \ket{0} = \sum_{j} c_j \ket{j}$, i.e., Eq. (\ref{eq-psi_UQU}).
This is the bit string for which we assume $\theta_{\rm ref} = 0$; the idea is that since physically 
no component in the  $\ket{j_{ref}}$ direction exists, $c_{ref}=0$, the phase should be irrelevant, thus $\theta_{ref} = 0$. 
By measuring the $m_{j_{\rm ref}} = \vert \bra{0_a \, j_{\rm ref}} \ket{\Psi^\prime}\vert^2$, 
and using the amplitude $\vert c_j \vert$,
in the limit of $\gamma \rightarrow 0^+$, 
the desired $c^{(i)}_j$ is computed.

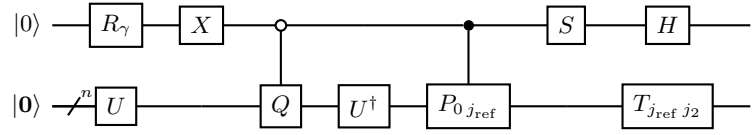
\begin{figure*}[t]
    \centering
    \begin{quantikz}
        \\
        \lstick{$\ket{0}$} 
        & \gate{R_{\gamma} } 
        & \gate{X} 
        & \octrl{1} 
        & \qw
        & \ctrl{1} 
        & \gate{S}  
        & \gate{H}  
        & \qw \\
        %%%%
        \lstick{$\ket{\textbf{0}}$} 
        & \gate{U} \qwbundle{n}
        & \qw
        & \gate{Q} 
        & \gate{U^\dagger} 
        & \gate{P_{0 \, j_{\rm ref}}}
        & \qw
        & \gate{T_{j_{\rm ref} \,  j_2}} 
        & \qw
        \\
    \end{quantikz}
    \caption{Quantum ciruit to measure real part $c^{(r)}_j$ (line 12 in Fig. \ref{fig:MQITE}).}
    \label{fig:mesurment-phase-real}
\end{figure*}

\emph{Real Part}.-- The quantum circuit to compute $c_j^{(r)}$
follows the same procedure but with an additional phase  gate $S$, as shown in the quantum circuit of Fig. \ref{fig:mesurment-phase-real}.

%%%%%
%%%%% fig 9
%%%%%
%%%%%%%%%%%%%%%%%%%%%%%%%%%%%%%%
%%%%%%%%%%%%%%%%%%%%%%%%%%%%%%%%

\section{An alternative derivation for gate parameters, Eqs. (\ref{eq-y-real}) and (\ref{eq-y-imag})}
\label{sec:Derivation2}
Here we provide a derivation 
for Eqs. (\ref{eq-y-real}) and (\ref{eq-y-imag})
based on the approach taken in variational quantum algorithms such as in \cite{Motta2019DeterminingEvolution}.

The basic idea is to minimize the difference between 
the incrementally propagated state via ITE and 
the state via the unitary equivalent such as QITE. 
From Eq. (\ref{eq-mqite-idea}), this is the difference 
between 
$$
e^{-\delta \, Q} U_{\tau}\ket{\textbf{0}} / \vert \vert 
e^{-\delta \, Q}\, U_{\tau}\ket{\textbf{0}} \vert \vert
$$
and 
$$
U_{\tau}\, e^{ i \sum_j  y^{(r)}_j \, P_j^{(r)} + y^{(i)}_j \,  P_j^{(i)}} \ket{\textbf{0}}.
$$ 
Up first order in $\delta$, assuming small parameters $y_j^{(r/i)} \sim {\cal O}(\delta)$, this difference results in the cost function
\begin{eqnarray}
f(\textbf{y}) = \left| \left| 
\sum_{j} y_j \, U_{\tau} \ket{j}
\,\, - \,\,
\delta \, 
\left[
Q - \langle Q \rangle
\right]\, 
\ket{\psi_\tau} 
\right| \right|.
\label{eq:cost-function}
\end{eqnarray}
Here, 
$\langle Q \rangle = \bra{\psi_{\tau}} Q \ket{\psi_{\tau}}$ and $\ket{\psi_{\tau}} = U_{\tau} \ket{\textbf{0}}$,
$y=y^{(r)} + iy^{(i)}$,
where $\vert \vert  \ket{\cdot} \vert \vert $ stands for $\bra{\cdot} \ket{\cdot}$.

After expanding the RHS of Eq. (\ref{eq:cost-function}) and finding the minimum of the cost $f$ by setting $\partial f / \partial y_j^{(r/i)} = 0$,
one 
arrives at Eqs. (\ref{eq-y-real}) and (\ref{eq-y-imag}) effortlessly.

The above procedure shows that, in contrast to QITE and other variational approach to implement ITE on a quantum computer, the formalism in this paper does not require solving any differential equation on a classical computer and the parameters can be obtained straightforwardly.

%%%%%%%%%%%%%%%%%%%%%%%%%%%%%%%%%%%%%%%%%%%%%%%%%%%%%
%%%%%%%%%%%%%%%%%%%%%%%%%%%%%%%%%%%%%%%%%%%%%%%%%%%%%
%%%%%%%%%%%%%%%%%%%%%%%%%%%%%%%%%%%%%%%%%%%%%%%%%%%%%
%%%%%%%%%
%%%%%%%%%
%%%%%%%%%.  list Qs and ws in nuc problem
%%%%%%%%%
%%%%%%%%%
%%%%%%%%%%%%%%%%%%%%%%%%%%%%%%%%%%%%%%%%%%%%%%%%%%%%%
%%%%%%%%%%%%%%%%%%%%%%%%%%%%%%%%%%%%%%%%%%%%%%%%%%%%%
%%%%%%%%%%%%%%%%%%%%%%%%%%%%%%%%%%%%%%%%%%%%%%%%%%%%%
\section{List of Pauli String Operators}
\label{sec:nuc-list-of-Qs}

The nuclear problem studied in the main text is initially in second quantization form. 
It is expressed in terms of Pauli strings using Jordan Wigner transformation \cite{jordan1993}. Below, the Pauli strings and associated weights are tabulated.
%\begin{center*}[t]
\begin{table*}[h]
\centering
\setlength{\tabcolsep}{10pt}
\begin{tabular}{c | cc | cc | cc}
\toprule
{} &         w &       Q &         w &       Q &         w &       Q \\
\hline
\midrule
0  & -0.446591 &  YXYZZX &  0.101688 &  YXXYII &  0.108894 &  XZXIXX \\
1  & -0.446591 &  YXXZZY & -0.101688 &  YYYYII & -0.108894 &  XZYIXY \\
2  & -0.446591 &  YYYZZY & -0.101688 &  XXXXII & -0.217787 &  YZZYZI \\
3  &  0.446591 &  YYXZZX &  0.101688 &  XYYXII & -0.217787 &  XZZXZI \\
4  &  0.446591 &  XXYZZY & -0.101688 &  XXYYII & -0.101688 &  YXIIYX \\
5  & -0.446591 &  XXXZZX & -0.101688 &  XYXYII & -0.101688 &  YYIIXX \\
6  & -0.446591 &  XYYZZX &  -0.18861 &  YZXYXI &  0.101688 &  YXIIXY \\
7  & -0.446591 &  XYXZZY &   0.18861 &  YZYYYI & -0.101688 &  YYIIYY \\
8  &  0.329172 &  YXIYXI &   0.18861 &  YZXXYI & -0.101688 &  XXIIXX \\
9  &   0.56401 &  YXIXYI &   0.18861 &  YZYXXI &  0.101688 &  XYIIYX \\
10 &  0.329172 &  YYIYYI &   0.18861 &  XZXYYI & -0.101688 &  XXIIYY \\
11 &  -0.56401 &  YYIXXI &   0.18861 &  XZYYXI & -0.101688 &  XYIIXY \\
12 &  -0.56401 &  XXIYYI &   0.18861 &  XZXXXI & -0.217787 &  YZZYIZ \\
13 &  0.329172 &  XXIXXI &  -0.18861 &  XZYXYI & -0.217787 &  XZZXIZ \\
14 &   0.56401 &  XYIYXI & -0.108894 &  YZXIYX &  0.217787 &  IYIZYI \\
15 &  0.329172 &  XYIXYI &  0.108894 &  YZYIYY &  0.217787 &  IXIZXI \\
16 &  0.435575 &  YZIYII &  0.108894 &  YZXIXY & -0.108894 &  IYXYZX \\
17 &  0.435575 &  XZIXII &  0.108894 &  YZYIXX &  0.108894 &  IYYYZY \\
18 & -0.101688 &  YXYXII &  0.108894 &  XZXIYY &  0.108894 &  IYXXZY \\
19 & -0.101688 &  YYXXII &  0.108894 &  XZYIYX &  0.108894 &  IYYXZX \\
\bottomrule
\end{tabular}
\caption{
Part 1 of the $84$ terms in the Hamiltonian used in the main paper for the nuclear physics application numerical example. 
}
\end{table*} 
%\end{center*}

%\begin{center*}
\begin{table*}[h]
\centering
\setlength{\tabcolsep}{10pt}
\begin{tabular}{c | cc | cc}
\toprule
{} &         w &       Q &         w &       Q \\
\midrule
0  &  0.108894 &  IXXYZY &  0.213531 &  IIXXXX \\
1  &  0.108894 &  IXYYZX & -0.213531 &  IIXYYX \\
2  &  0.108894 &  IXXXZX &  0.213531 &  IIXXYY \\
3  & -0.108894 &  IXYXZY &  0.213531 &  IIXYXY \\
4  &  0.217787 &  IYZIYI &       &   \\
5  &  0.217787 &  IXZIXI &       &   \\
6  &  -0.18861 &  IYZXYX &      &    \\
7  &   0.18861 &  IYZYYY &      &    \\
8  &   0.18861 &  IYZXXY &      &    \\
9  &   0.18861 &  IYZYXX &       &   \\
10 &   0.18861 &  IXZXYY &     &     \\
11 &   0.18861 &  IXZYYX &      &    \\
12 &   0.18861 &  IXZXXX &     &     \\
13 &  -0.18861 &  IXZYXY &     &     \\
14 & -0.435575 &  IYZZYZ &       &   \\
15 & -0.435575 &  IXZZXZ &      &    \\
16 &  0.213531 &  IIYXYX &       &   \\
17 &  0.213531 &  IIYYXX &       &   \\
18 & -0.213531 &  IIYXXY &      &    \\
19 &  0.213531 &  IIYYYY &      &    \\
\bottomrule
\end{tabular}

\caption{
Part 2 of the terms in the Hamiltonian used in the main paper for the nuclear physics application numerical example. The Hamiltonian is expressed by Jordan Wigner transformation in terms of Pauli string operators.
}
\end{table*} 
%\end{center*}

% \clearpage
%-----------------------------------------------------------

\end{document}